\documentclass[preprint,review,12pt]{elsarticle}

\usepackage{graphics}
\usepackage{graphicx}

\usepackage{epsfig}

\usepackage{amssymb}
\usepackage{amsthm}

\newtheorem*{remark}{Remark}

\usepackage{lineno}
\usepackage{color}

\begin{document}

\begin{frontmatter}

 

\title{
Turbulent energy cascade through equivalence of Euler and Lagrange motion descriptions and bifurcation rates 
}
 

\author{Nicola de Divitiis}

\address{"La Sapienza" University, Dipartimento di Ingegneria Meccanica e Aerospaziale, 
Via Eudossiana, 18, 00184 Rome, Italy, \\
phone: +39--0644585268, \ \ fax: +39--0644585750, \\ 
e-mail: n.dedivitiis@gmail.com, \ \  nicola.dedivitiis@uniroma1.it}

\begin{abstract}
This work analyses the homogeneous isotropic turbulence by means of the equivalence between Euler and Lagrange representations of motion, adopting the bifurcation rates associated with Navier--Stokes and kinematic equations, and an appropriate hypothesis of fully developed chaos.
The equivalence of these motion descriptions allows to show that kinetic and thermal energy cascade arise both from the convective term of Liouville equation. Accordingly, these  phenomena, of nondiffusive nature, correspond to a transport in physical space linked to the trajectories divergence.
Both the bifurcation rates are properly defined, where the kinematic bifurcation rate is shown to be much greater than Navier--Stokes bifurcation rate. This justifies the proposed hypothesis of fully developed chaos where velocity field and particles trajectories fluctuations are statistically uncorrelated. 
Thereafter, a specific ergodic property is presented, which relates the statistics of fluid displacement to that of velocity and temperature fields. 
A detailed analysis of separation rate is proposed which studies the statistics of radial velocity component along the material separation vector. Based on previous elements, the closure formulas of von K\'arm\'an--Howarth and Corrsin equations are finally achieved. These closures, of nondiffusive kind, represent a propagation phenomenon, and coincide with those just presented by the author in previous works, corroborating the results of these latter. 
\end{abstract}

\begin{keyword}
Turbulence, Lagrange description, Euler description, Bifurcation rate, Liouville theorem.
\end{keyword}

\end{frontmatter}

\newcommand{\no}{\noindent}
\newcommand{\be}{\begin{equation}}
\newcommand{\ee}{\end{equation}}
\newcommand{\bea}{\begin{eqnarray}}
\newcommand{\eea}{\end{eqnarray}}
\newcommand{\bc}{\begin{center}}
\newcommand{\ec}{\end{center}}

\newcommand{\calr}{{\cal R}}
\newcommand{\calv}{{\cal V}}

\newcommand{\bff}{\mbox{\boldmath $f$}}
\newcommand{\bfg}{\mbox{\boldmath $g$}}
\newcommand{\bfh}{\mbox{\boldmath $h$}}
\newcommand{\bfi}{\mbox{\boldmath $i$}}
\newcommand{\bfm}{\mbox{\boldmath $m$}}
\newcommand{\bfp}{\mbox{\boldmath $p$}}
\newcommand{\bfr}{\mbox{\boldmath $r$}}
\newcommand{\bfu}{\mbox{\boldmath $u$}}
\newcommand{\bfv}{\mbox{\boldmath $v$}}
\newcommand{\bfx}{\mbox{\boldmath $x$}}
\newcommand{\bfy}{\mbox{\boldmath $y$}}
\newcommand{\bfw}{\mbox{\boldmath $w$}}
\newcommand{\bfk}{\mbox{\boldmath $\kappa$}}

\newcommand{\bfA}{\mbox{\boldmath $A$}}
\newcommand{\bfD}{\mbox{\boldmath $D$}}
\newcommand{\bfI}{\mbox{\boldmath $I$}}
\newcommand{\bfL}{\mbox{\boldmath $L$}}
\newcommand{\bfM}
{\mbox{\boldmath $M$}}
\newcommand{\bfS}{\mbox{\boldmath $S$}}
\newcommand{\bfT}{\mbox{\boldmath $T$}}
\newcommand{\bfU}{\mbox{\boldmath $U$}}
\newcommand{\bfX}{\mbox{\boldmath $X$}}
\newcommand{\bfY}{\mbox{\boldmath $Y$}}
\newcommand{\bfK}{\mbox{\boldmatthe average of $u_\xi u_\xi^*/u^2$h $K$}}

\newcommand{\bfeta}{\mbox{\boldmath $\eta$}}
\newcommand{\bfiota}{\mbox{\boldmath $\iota$}}
\newcommand{\bfrho}{\mbox{\boldmath $\rho$}}
\newcommand{\bfchi}{\mbox{\boldmath $\chi$}}
\newcommand{\bfphi}{\mbox{\boldmath $\phi$}}
\newcommand{\bfPhi}{\mbox{\boldmath $\Phi$}}
\newcommand{\bflambda}{\mbox{\boldmath $\lambda$}}
\newcommand{\bfxi}{\mbox{\boldmath $\xi$}}
\newcommand{\bfLambda}{\mbox{\boldmath $\Lambda$}}
\newcommand{\bfPsi}{\mbox{\boldmath $\Psi$}}
\newcommand{\bfomega}{\mbox{\boldmath $\omega$}}
\newcommand{\bfOmega}{\mbox{\boldmath $\Omega$}}
\newcommand{\bfeps}{\mbox{\boldmath $\varepsilon$}}
\newcommand{\bfepsn}{\mbox{\boldmath $\epsilon$}}
\newcommand{\bfzeta}{\mbox{\boldmath $\zeta$}}
\newcommand{\bfkappa}{\mbox{\boldmath $\kappa$}}
\newcommand{\bfsigma}{\mbox{\boldmath $\sigma$}}
\newcommand{\bftau}{\mbox{\boldmath $\tau$}}
\newcommand{\itPsi}{\mbox{\it $\Psi$}}
\newcommand{\itPhi}{\mbox{\it $\Phi$}}

\newcommand{\bint}{\mbox{ \int{a}{b}} }
\newcommand{\ds}{\displaystyle}
\newcommand{\Sum}{\Large \sum}



\bigskip

\section{Introduction \label{intro}}

Classical studies \cite{Karman38, Corrsin_1, Corrsin_2} analyzed homogeneous isotropic turbulence by means of correlations evolution equations using the Eulerian (or spatial) description of fluid motion. There, velocity and temperature fields ensembles 
(Eulerian ensembles) were adopted to define correlations of velocity and temperature, without considering the Lagrangian standpoint i.e the fluid displacement evolution. Although these studies give correlations evolution equations, such works neither explain the energy cascade nor provide the closure formulas for the convective terms of correlation equations unless specific mathematical structures of such closures are assumed 
\cite{Hasselmann58, Millionshtchikov69, Oberlack93, Baev, Mellor84}. 
The reason for this could be due to the fact that fluid displacement and particles trajectories, which play important roles in turbulence, are not adequately expressed in Eulerian description, at least in the framework of correlations equations. 
Viceversa, such elements are expressly defined in Lagrange standpoint. 
However, the displacement of a mechanical system contributes to define the state of motion
of this latter, where the corresponding phase space is made by generalized coordinates and velocities (impulses). The reduction of fluid dynamics phase space to velocity and temperature fields spaces (Eulerian description), without considering fluid displacement (Lagrangian element), could mean that the continuum particles transport, although incorporated in the spatial standpoint, is not properly expressed for the purposes of turbulence description.
On the other hand,  Euler and Lagrange representations are equivalent view points 
if the fluid motion satisfies very general smoothness conditions \cite{Truesdell77}.
Therefore, the basic idea of this work is of using the equivalence between the two descriptions to study the energy cascade and a specific hypothesis of fully developed chaos which establishes the statistical independence of velocity field and fluid displacement. 
To justify such hypothesis, the concept of bifurcation rate is first introduced. Such bifurcation rate is the average frequency at which the bifurcations happen during the chaotic regimes. Specifically, this is the frequency at which the trajectories intersect $\Sigma_D$, the hypersurface of phase space where the system jacobian determinant vanishes. Of course, if the trajectories do not cross $\Sigma_D$, the system, although nonlinear, will not exhibit chaotic behavior. On the contrary, if trajectories continuously intersect $\Sigma_D$, the chaotic behavior is observed and the state variables fluctuate with a rapidity expressed by the bifurcation rate.
Specifically, the present analysis considers two kinds of bifurcation rates: the bifurcation rate associated with Navier--Stokes equations and that relative to fluid displacement equation.

Although many works were written regarding the homogeneous isotropic turbulence and the closures of correlation equations  
\cite{Hasselmann58, Millionshtchikov69, Oberlack93, Baev, Mellor84} \cite{George1, George2, Antonia,  Onufriev94, Grebenev05, Grebenev09, Antonia2013}, to the author's knowledge an analysis based on both the aspects of bifurcation rates and motion descriptions has not received due attention. Therefore, the aim of the this work is to study homogeneous isotropic turbulence by means of a proper hypothesis of fully developed chaos, using both Euler and Lagrange points of view and the bifurcation rate of nonlinear systems. 

This work first recall the fundamentals of continuum fluid kinematics useful for the present analysis \cite{Truesdell77}, thereafter, to formulate the Liouville theorem for both Eulerian and Lagrangian points of view, mass, momentum and heat equations are considered together to the displacement evolution equation. The obtained set of differential equations is first reduced to the symbolic form of operators. This allows to define Navier--Stokes bifurcations and kinematic bifurcations as in the case of ordinary differential equations, and leads to study different aspects of fully developed turbulence some of which ones are just analyzed by the author in previous works \cite{deDivitiis_1,  deDivitiis_4, deDivitiis_5, deDivitiis_8, deDivitiis_9}.
The novelty of the present work with respect to such articles consists in the following items: 

(i) The orders of magnitude of bifurcation rates are estimated by means of a specific analysis which uses the velocity correlation equation and exploits the properties of bifurcations effects on velocity fields and trajectories. The kinematic bifurcation rate is shown to be much greater than Navier--Stokes bifurcation rate and this justifies the proposed hypothesis of fully developed chaos following which velocity field (Eulerian element) and trajectories in physical space (Lagrangian element) are statistically uncorrelated in fully developed turbulence. 

(ii) Through the equivalence between Lagrange and Euler standpoints, kinetic and thermal energy cascade are shown to be arising from the same transport term of Liouville equation.

(iii) Based on (i) and statistical homogeneity, a specific ergodic property is proposed, which allows to relate the statistics of Lagrangian description to that of Eulerian point of view. 

(iv) A detailed mathematical analysis of separation rate based on the previous items is presented, which leads to estimate the range of longitudinal velocity component along the material separation vector and the corresponding statistics. 

The closures of von K\'arm\'an--Howarth equation and Corrsin equation are then achieved through the previous items. These closures, here derived using the Liouville theorem, coincide with those just obtained by the author in previous articles \cite{deDivitiis_1,  deDivitiis_4, deDivitiis_5, deDivitiis_8, deDivitiis_9}. These formulas, of nondiffusive type, represent a kind of correlations propagation phenomenon along the separation distance $r$, and allow to adequately describe the energy cascade, providing values of skewness of longitudinal velocity derivative and temperature derivative equal to -3/7 and -1/5, respectively.

\bigskip

\section{Background: Kinematics of Continuum Fluids \label{kinematics}}

To study some of the statistical properties of fully developed turbulence, the fundamentals of continuum fluid kinematics are first renewed according to the classical theoretical formulation \cite{Truesdell77}. 
Such fundamentals, regarding the motion representations, will be useful for present analysis. 

In particular, this background (a) remarks that the displacement effects, although incorporated in  Euler standpoint, could not be adequately expressed for the purposes of turbulence representation through correlations, and (b) provides plausible arguments that displacement fluctuations can be much more rapid than velocity and temperature fields in fully developed turbulence.
\begin{figure} 
	\centering
	\includegraphics[width=80mm,height=70mm]{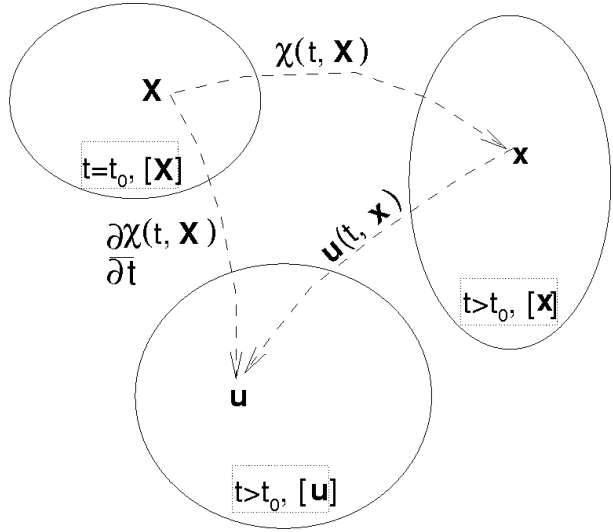}
\caption{Scheme of fluid displacement and velocity field.}
\label{figura_1}
\end{figure}

\begin{figure}
	\centering
	\includegraphics[width=80mm,height=70mm]{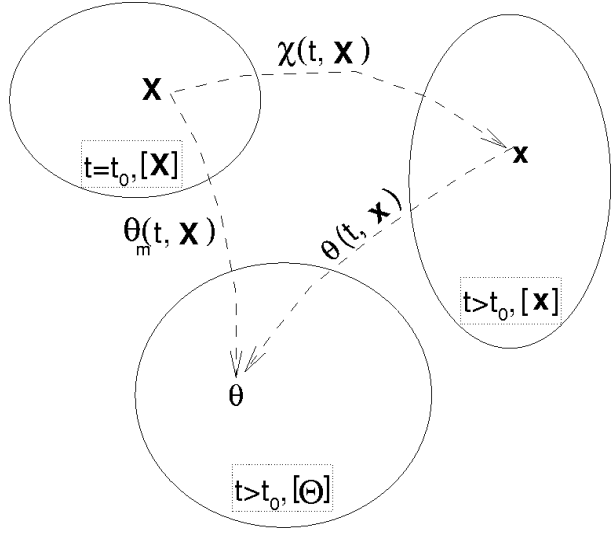}
\caption{Scheme of fluid displacement and temperature field.}
\label{figura_2}
\end{figure}

To analyze this, the following map is now considered (see schemes of Figs. \ref{figura_1} and \ref{figura_2}) \cite{Truesdell77}
\bea
\ds {\bfchi}(t, .): {\bf X} \rightarrow {\bfx}(t)
\eea
This expresses the referential motion representation which gives the displacement at current time 
$t$ $\ne$ $t_0$ of a fluid particle placed in $\bf X$ (referential positions) at $t$=$t_0$. 
Thus, $\bf X$ plays also the role of a label which uniquely identifies a particle that at $t$=$t_0$ is placed on $\bf X$. More in general, the referential configuration is the shape that the fluid occupies at $t$=$t_0$ or could occupy. Viceversa, $\bf X$ can be formally expressed in terms of $\bf x$, through the inverse map $\bfchi^{-1}$  \cite{Truesdell77} 
\bea
\begin{array}{l@{\hspace{-0.cm}}l}
\ds {\bf x} = {\bfchi}(t, {\bf X}) \\\\
\ds {\bf X} = {\bfchi}^{-1}(t, {\bf x}) 
\end{array}
\eea
The velocity of $\bf X$ is then defined as 
\bea
\ds \dot{\bfx} \equiv \dot{\bfchi} \equiv \frac{\partial \bfchi}{\partial t} (t, {\bf X})
\label{v material}
\eea
and the temperature of $\bf X$ is written as
\bea
\vartheta_m = \vartheta_m(t, {\bf X})
\label{t material}  
\eea
Equations (\ref{v material})--(\ref{t material}) give Lagrangian or referential representation
of fluid motion, being $\dot{\bfchi}(t, {\bf X})$ and $\vartheta_m(t, {\bf X})$ velocity and temperature variations along the trajectory of $\bf X$. 

Following the Eulerian view point, velocity and temperature fields, ${\bf u}(t, {\bf x})$ and 
$\vartheta(t, {\bf x})$, are defined according to the schemes of Figs. \ref{figura_1} and \ref{figura_2}, by means of $\dot{\bfchi}$, $\vartheta_m$ and the map $\bfchi^{-1}$
\bea
\begin{array}{l@{\hspace{-0.cm}}l}
\ds {\bf u}(t, {\bf x})= \frac{\partial{\bfchi}}{\partial t}\left( t, {\bfchi}^{-1}(t, {\bf x}) \right)
\\\\
\ds \vartheta(t, {\bf x})= \vartheta_m\left(t, {\bfchi}^{-1}(t, {\bf x}) \right)  
\end{array}
\label{fields}
\eea
It is worth to remark that such Eulerian fields are not directly defined as functions of $t$ and 
$\bf x$. These are defined starting from Eqs. (\ref{v material})--(\ref{t material}), as restrictions of $\dot{\bfchi}(t, {\bf X})$ and $\vartheta_m(t, {\bf X})$ on the motion ${\bfchi}(t, {\bf X})$ \cite{Truesdell77}.
Although the two representations are equivalent with each other, and the Eulerian description incorporates the informations of fluid displacement through Eq. (\ref{fields}), unlike the referential standpoint, the Euler point of view does not express such informations in explicit way as ${\bfchi}(t, {\bf X})$ is implicitly incorporated in ${\bf u}(t, {\bf x})$ and $\vartheta(t, {\bf x})$. 
As the result, the transport effects of $\bfchi$, very important in turbulence,
could not be adequately represented through the sole knowledge of Eulerian fields, 
at least for the turbulence description in terms of correlations. 
Such effects are calculated through velocity and temperature fields as  
\bea
\begin{array}{l@{\hspace{-0.cm}}l}
\ds \dot{\bfchi}(t, {\bf X})= {\bf u} \left( t, {\bfchi}(t, {\bf X}) \right), \\\\
\ds \vartheta_m(t, {\bf X})= \vartheta\left(t, {\bfchi}(t, {\bf X}) \right)
\label{t fields}
\end{array}
\eea
Hence, Lagrangian and Eulerian fields are defined in the functions spaces $\left\lbrace \dot{\bfchi}(t, {\bf X}) \right\rbrace  \times  \left\lbrace \vartheta_m(t, {\bf X}) \right\rbrace$ and $\left\lbrace {\bf u}(t, {\bf x}) \right\rbrace  \times  \left\lbrace {\vartheta}(t, {\bf x}) \right\rbrace$, respectively.

At this stage of present analysis, the following should be noted
\begin{remark}
As the result of previous definitions, for an assigned motion, there are many different referential descriptions, equally valid, depending on the referential displacement, whereas the corresponding Eulerian fields ${\bf u}(t, {\bf x})$ and $\vartheta(t, {\bf x})$ are unique \cite{Truesdell77}. 
\end{remark}
\begin{remark}
Due to huge level of chaos and mixing observed in developed turbulence, according to 
Eqs. (\ref{fields})--(\ref{t fields}), the various fields
${\bfchi}(t, {\bf X})$ are expected to be much faster than  ${\bf u}(t, {\bf x})$. 
If ${\bf u}(t, {\bf x})$ is a slow growing function of $t$, then ${\bfchi}(t, {\bf X})$ could exponentially vary with respect to time. 
Accordingly, also ${\vartheta}(t, {\bf x})$ can be much 
slower than ${\bfchi}(t, {\bf X})$ and ${\vartheta}_m(t, {\bf X})$. 
\label{rem1}
\end{remark}

For sake of our convenience, the ratio between elemental volumes of functions spaces $\left\lbrace \dot{\bfchi}(t, {\bf X}) \right\rbrace  \times  
\left\lbrace \vartheta_m(t, {\bf X}) \right\rbrace$ and $\left\lbrace {\bf u}(t, {\bf x}) \right\rbrace  \times  \left\lbrace {\vartheta}(t, {\bf x}) \right\rbrace$ is now calculated 
when $\bfchi$ is assigned and ${\bf x} = \bfchi(t, {\bf X})$. 
This ratio is the absolute value of the determinant of the jacobian $\bf J$
\bea
\ds {\bf J} = \left( \frac{\delta\left( {\bf u}, \vartheta \right)  }{\delta \left(\dot{\bfchi}, \vartheta_m\right) }\right)_{\bfchi} \equiv {\bf I},
\label{jac}
\eea
where $\delta ()/ \delta ()$  stands for functional derivatives defined in the corresponding  functions spaces, and the subscript $\bfchi$ here indicates that $\bf J$ is calculated for $\bfchi$ given and ${\bf x} = \bfchi(t, {\bf X})$. 
According to Eq. (\ref{fields}) or (\ref{t fields}), $\bf J$ identically equals the identity operator $\bf I$, therefore its determinant is
\bea
\ds \det {\bf J} = \det \left( \frac{\delta\left( {\bf u}, \vartheta \right)  }{\delta \left(\dot{\bfchi}, \vartheta_m\right) }\right)_{\bfchi} \equiv 1,
\label{det jac}
\eea
More in detail, the following functional derivatives, also computed in the same conditions
\bea
\ds  \left( \frac{\delta {\bf u} }{\delta \dot{\bfchi}}\right)_{\bfchi} \equiv {\bf I}, \ \ \ \
\ds  \left( \frac{\delta \vartheta }{\delta {\vartheta_m}}\right)_{\bfchi} \equiv {\bf I}
\label{jac 1}
\eea
correspond to identity operators, each defined in the relative function space. 

The considerations about the fluid temperature apply also to any passive scalar.

Such background gives the starting elements of present analysis regarding 
some of the statistical properties of developed turbulence, 
useful for this study.

\bigskip

\section{Background: Evolution equations, phase space and motion descriptions \label{Background}}

The aim of this section is to define the set of the evolution equations of fluid state variables and
the corresponding phase spaces. To this end, these equations are expressed in infinite domain for both the representations, where the fluid motion is expressed with respect to an assigned inertial frame $\cal R$.

The Eulerian representation describes the motion by means of fluid properties functions of $t$ and $\bf x$. In this case,  mass, momentum equations (Navier--Stokes equations) and heat equation are 
\bea
\begin{array}{l@{\hspace{-0.cm}}l}
\ds \nabla_{\bf x} \cdot {\bf u} =0, \\\\
\ds \dot{\bf u} \equiv \frac{\partial {\bf u}}{\partial t} =
-  \nabla_{\bf x} {\bf u} \ {\bf u} - \frac{\nabla_{\bf x}p}{\rho}  + \nu \nabla_{\bf x}^2 {\bf u}
\label{NS_eq Euler}
\end{array}
\eea
\bea
\begin{array}{l@{\hspace{-0.cm}}l}
\ds \dot{\vartheta} \equiv \frac{\partial \vartheta }{\partial t} =
-{\bf u} \cdot \nabla_{\bf x} \vartheta + \kappa \nabla_{\bf x}^2 \vartheta,
\label{T_eq Euler}
\end{array}
\eea
Thus, ${\bf u}(t, {\bf x})$ and $\vartheta(t, {\bf x})$ are the fluid state variables associated with this description. According to Eq. (\ref{fields}), ${\bf u}(t, {\bf x})$ and 
$\vartheta(t, {\bf x})$ are linked to  ${\bfchi}(t, {\bf X})$, where this latter does not influence Eqs. (\ref{NS_eq Euler}) and (\ref{T_eq Euler}), thus ${\bfchi}(t, {\bf X})$ is not a state variable and its evolution is implicitly included in Eqs. (\ref{NS_eq Euler}). 

For Lagrangian point of view, material properties are calculated by tracking fluid particles during the motion. In this case, the evolution equations can be written in the following way
\bea
\begin{array}{l@{\hspace{-0.cm}}l}
\ds \mbox{tr} \left( \left( \nabla_{\bf X} \dot{\bfchi} \right)  \left(  \nabla_{\bf X}{\bfchi} \right)^{-1} \right)=0, \\\\ 
\ds \dot{\bfchi}^{(1)}\equiv \ddot{\bfchi}\equiv \frac{D{\bf u}}{Dt} \equiv \frac{\partial {\bf u}}{\partial t} + \nabla_{\bf x} {\bf u} \ {\bf u} = 
\frac{1}{\rho} \ \nabla_{\bf X} \cdot {\bf T}
\label{NS_eq Lagrange}
\end{array}
\eea
\bea
\begin{array}{l@{\hspace{-0.cm}}l}
\ds \dot{\vartheta}_m \equiv \frac{D{\vartheta}}{Dt} \equiv \frac{\partial \vartheta }{\partial t} + {\bf u} \cdot \nabla_{\bf x} \vartheta = 
\frac{1}{\rho C_p} \  \nabla_{\bf X} \cdot {\bf q} 
\label{T_eq Lagrange}
\end{array}
\eea
\bea
\ds \dot{\bfchi} \equiv {\bf u}\left( t, {\bfchi} \right) = {\bfchi}^{(1)} 
\label{chi}
\eea
where $\bf T$ and $\bf q$ are stress tensor and heat flux density respectively. 
In this formulation, Eq. (\ref{chi}) is the evolution equation of fluid placement, 
thus $\bfchi$ is a state variable of Lagrangian point of view.

The state variables for Lagrange standpoint are $\dot{\bfchi}(t, {\bf X})$, ${\bfchi}(t, {\bf X})$ and $\vartheta_m(t, {\bf X})$, whereas those associated with Eulerian representation are given by ${\bf u}(t, {\bf x})$ and $\vartheta(t, {\bf x})$, thus the corresponding phase spaces are $\cal E$ and $\cal L$, being
\bea
\begin{array}{l@{\hspace{-0.cm}}l}
\ds {\cal E} = \left\lbrace {\bf u}(t, {\bf x}) \right\rbrace  \times  \left\lbrace {\vartheta}(t, {\bf x}) \right\rbrace, \\\\ 
\ds {\cal L}= \left\lbrace {\bfchi}(t, {\bf X}) \right\rbrace \times \left\lbrace \dot{\bfchi}(t, {\bf X}) \right\rbrace  \times  
\left\lbrace \vartheta_m(t, {\bf X}) \right\rbrace
 \end{array}
\label{phase spaces}
\eea
Here, for sake of convenience, $\cal E$ and ${\cal X} \equiv \lbrace {\bfchi}(t, {\bf X}) \rbrace$ are called Eulerian and Lagrangian sets respectively, and the corresponding sets of solutions are Euler and Lagrange ensembles.

Observe that, according to Eq. (\ref{phase spaces}) and to remarks of previous section, there is no
one--to--one correspondence between the points of $\cal L$ and those of $\cal E$. One point of $\cal E$
corresponds to infinite points of $\cal L$ which are all equally valid for the motion description. 
This determines a high level of uncertainty of displacement effects which may be unacceptable for
the purposes of turbulence description in terms of correlations.
Here, in order to obtain a one--to--one correspondence between $\cal E$ and $\cal L$ which formally preserves all the informations of fluid displacement, the phase space $\cal E$ is suitably augmented with  $\cal X$, introducing $\hat{\bfchi}$, the fluid displacement frozen at current time, whose evolution equation is  
\bea
\ds \dot{\hat{\bfchi}} = 0
\label{dot hat chi}
\eea

For what concerns the various quantities which appear in Eqs. (\ref{NS_eq Euler}) and (\ref{T_eq Euler}), $p$=$p(t, {\bf x})$, $\nu$ and $\kappa$=$k/ \rho C_p$ are pressure, kinematic viscosity and thermal diffusivity, respectively,  being $\rho=$const, $k$ and $C_p$ density, fluid thermal conductivity and specific heat at constant pressure.
Here, $\nu$ and $\kappa$ are supposed to be independent of temperature, thus for both
the representations, momentum equations are autonomous with respect to heat equation,
whereas the solutions of Eqs. (\ref{T_eq Euler}) and (\ref{T_eq Lagrange}) will depend on 
Eqs. (\ref{NS_eq Euler}) and (\ref{NS_eq Lagrange}), respectively.
$p$ is first eliminated through continuity equation, and the momentum equation is given in terms of velocity for both the representations.
Specifically, $p$ is reduced to be a functional of ${\bf u}(t, {\bf x})$ through continuity equation. This makes the momentum equations integro--differential equations, where $p$ exerts a nonlocal effect \cite{Tsinober2009} on  fluid motion.

\bigskip

\section{Bifurcations of Navier--Stokes equations}

The bifurcations of evolution equations are now defined as in the case of ordinary differential equations. As Eulerian and Lagrangian descriptions are equivalent representations, the bifurcations are here considered only for Eqs. (\ref{NS_eq Euler})--(\ref{T_eq Euler}). 
To this end, in line with Refs. \cite{deDivitiis_5, deDivitiis_8}, such equations are first reduced to the symbolic form of operators as
\bea
\begin{array}{l@{\hspace{-0.cm}}l}
 \dot{\bf u} =  {\bf N}({\bf u} ; \nu),
\end{array}
\label{NS_op}
\eea
\bea
\begin{array}{l@{\hspace{-0.cm}}l}
\dot{\vartheta} =  {\bf M}({\bf u}, \vartheta ; \kappa)
\end{array}
\label{T_op}
\eea
${\bf N}$ is a quadratic operator including, among the other terms, the integral nonlinear operator which gives $\nabla_{\bf x} p$ as functional of 
${\bf u}$, i.e. 
\bea
\begin{array}{l@{\hspace{-0.cm}}l}
\ds {\bf N} = {\bf L} {\bf u} + \frac{1}{2} {\bf C} {\bf u} \ {\bf u}
\end{array}
\label{Nw}
\eea
First and second terms of Eq. (\ref{Nw}) formally represent viscous forces and the contribution of both inertia and pressure forces, respectively, being $\bf L$ and $\bf C$ proper operators. As for Eq. (\ref{T_op}), it represents the evolution equation of $\vartheta$, where $\bf M$ linearly acts on $\vartheta$, and depends on ${\bf u}$ and $\kappa$.

According to Ref. \cite{Ruelle71}, Eqs (\ref{NS_op}) and (\ref{T_op}) are here analyzed supposing that the infinite dimensional phase space ${\cal E}$ (thus also $\left\lbrace \bf u \right\rbrace$ and  $\left\lbrace \vartheta \right\rbrace$) can be dealt with as a finite-–dimensional manifold. This method of analysis, considered to be valid in the limits of the formulation of Ref. \cite{Ruelle71}, allows to formally apply the classical bifurcation theory of ordinary differential equations \cite{Ruelle71, Eckmann81,  Guckenheimer90} to Eqs. (\ref{NS_op}) and (\ref{T_op}).

As $\bf M$ is a linear operator, transition and turbulence are caused by the bifurcations of Eq. (\ref{NS_op}), where $\nu^{-1}$ plays the role of control parameter. 
Such bifurcations occur in the points of $\left\lbrace \bf u \right\rbrace$ where the Jacobian ${\nabla_{\bf u} {\bf N}}$ exhibits at least an eigenvalue with zero real part (NS--bifurcations), and this occurs when
\bea
\ds \Sigma_u: \ {\cal D}_{NS} \equiv \det (\nabla_{\bf u} {\bf N}) =0.
\label{det NS}
\eea
where $\bf A \equiv {\nabla_{\bf u} {\bf N}}$ is linear with respect to $\bf u$, i.e.
\bea
\begin{array}{l@{\hspace{-0.cm}}l}
\ds {\bf A} \equiv \nabla_{\bf u} {\bf N} = {\bf L} + {\bf C_s} {\bf u}
\end{array}
\label{grad N}
\eea
and $\bf C_s \ u$ is the symmetrical part of $\bf C \ u$, properly defined.
The form (\ref{det NS}) is a kind of secular equation, representing a hypersurface $\Sigma_u \subset \left\lbrace \bf u \right\rbrace$, where ${\cal D}_{N S}$ is an infinite Taylor series of $\bf u$ expressed accounting for Eq. (\ref{grad N}). As the result, $\Sigma_u$ is expected to be a smooth hypersurface which moves in velocity field set. 

\bigskip

\subsection{Estimating Navier--Stokes bifurcation rate}

During the turbulent motion, the continuous vanishing of ${\cal D}_{N S}$ corresponds to velocity field fluctuations whose rapidly directly arises from the rate at which the Navier--Stokes bifurcations happen.
This rate,  defined by
\bea
\begin{array}{l@{\hspace{-0.cm}}l}
\ds S_{NS} = \lim_{T \rightarrow \infty} \frac{1}{T} \int_0^T \delta({\cal D}_{NS}) \vert \frac{d {\cal D}_{NS}}{dt} \vert \ dt
\end{array}
\label{NS rate}
\eea
is the average frequency at which the phase trajectories  intersect $\Sigma_u \subset \left\lbrace {\bf u}\right\rbrace$, where 
\bea
\begin{array}{l@{\hspace{-0.cm}}l}
\ds \frac{d {\cal D}_{NS}}{dt} = {\cal D}_{NS} \ \mbox{tr}\left( 
{\bf A}^{-1} 
\frac{d 
{\bf A}}{dt} \right), \\\\
\ds  \frac{d {\bf A}}{dt} = \nabla_{\bf u}  {\bf A} \ \dot{\bf u}
\end{array}
\label{app}
\eea
in which $\mbox{tr} \left( \circ \right)$ stands for the trace of $\circ$. To estimate $S_{NS}$ in homogeneous isotropic turbulence, observe that, the order of magnitude of the fluctuations of $\bf u$ can be written in terms of $S_{NS}$ as follows
\bea
\ds {\bf u} \cdot{\bf u} \sim \frac{{\bf u} \cdot \dot{\bf u}}{S_{N S}}
\eea
and this leads to estimate $S_{NS}$ as follows
\bea
\begin{array}{l@{\hspace{-0.cm}}l}
\ds S_{NS} \backsim \vert \frac{d}{dt} \ln u \vert, 
\end{array}
\eea
being $u = \sqrt{\langle {\bf u} \cdot {\bf u} \rangle_{\cal E}/3}$, where 
$\langle \circ \rangle_{\cal E}$ denotes the average of $\circ$ computed over the Eulerian ensamble. 
On the other hand, the evolution equation of velocity standard deviation in homogeneous isotropic turbulence \cite{Karman38}, allows $S_{NS}$ to be expressed in function of kinematic viscosity and velocity correlation length 
\bea
\begin{array}{l@{\hspace{-0.cm}}l}
\ds S_{NS}  \backsim \frac{\nu}{\lambda_T^2 } = \frac{u}{\lambda_T} \frac{1}{R_T}, \\\\
\ds R_T = \frac{u \lambda_T}{\nu}
\end{array}
\label{P S_NS}
\eea
where $\lambda_T$ and $R_T$ are, respectively, velocity correlation length (Taylor scale),
and Taylor scale Reynolds number.
\begin{remark}
We conclude this section observing that the NS--bifurcations regard the entire velocity field: 
these produce a global effect on motion which influences all  points of fluid domain. 
\end{remark}

\bigskip

\section{Kinematic analysis}

The Navier--Stokes bifurcations produce velocity and temperature fields doubling in the sense that all the properties associated with these fields are doubled, with particular reference to their characteristic scales $\ell_q$ and times $\tau_q$, $q=$ 1, 2,... \cite{deDivitiis_8} where $q$ stands for the number of encountered NS bifurcations when $\nu \rightarrow$0 plays the role of control parameter of Eq. (\ref{NS_op}). In detail, consider now a velocity field at a given time. When $\nu$ decreases, after the occurrence of several bifurcations, the velocity field will be represented by a function of the kind
\bea
\begin{array}{l@{\hspace{-0.cm}}l}
\ds {\bf u}\left(t, {\bf x} \right) = {\bf u}\left( \frac{t}{\tau_1}, \frac{t}{\tau_2},..., \frac{t}{\tau_q}; \ \frac{\bf x}{\ell_1}, \frac{\bf x}{\ell_2},...,\frac{\bf x}{\ell_q} \right), 
\end{array}
\label{u l}
\eea
where, according to the theory \cite{Ruelle71, Feigenbaum78, Pomeau80, Eckmann81}, the turbulence starts when $q \gtrsim 3, 4$, thereafter $q$ diverges while $\ell_k$ and $\tau_k$ are continuously distributed.

In fully developed turbulence, the minimum of such scales, say
$\ell_{min} = \mbox{min}\left\lbrace \ell_k, k= 1, 2,... \right\rbrace \equiv \ell_m$, is identified with
the condition that inertia and viscosity forces are locally balanced with each other, i.e.
\bea 
\begin{array}{l@{\hspace{-0.cm}}l}
\ds \frac{u_k \ell_{min}}{\nu} \approx 1, \\\\
\ds \ell_{min} = u_k \ \tau_m, \\\\
\ds \lambda_T = u \ \tau_m.
\end{array}
\label{small scale}
\eea
Combining Eqs. (\ref{small scale}) and eliminating $\tau_m$, one obtains the relation
\bea 
\begin{array}{l@{\hspace{-0.cm}}l}
\ds \frac{\lambda_T}{\ell_{min}} \approx \frac{u}{u_k} \approx \sqrt{R_T}
\end{array}
\label{Taylor kolmogorov}
\eea
which establishes that $\ell_{min}$ and $u_k$ identify, respectively, Kolmogorov scale and the corresponding velocity.
Accordingly, the trajectory of an arbitrary fluid particle, $\bf X$, calculated following Eq. (\ref{u l})
\bea
\begin{array}{l@{\hspace{-0.cm}}l}
\ds \dot{\bfchi}\left(t, {\bf X} \right) = {\bf u}\left( \frac{t}{\tau_1}, \frac{t}{\tau_2},..., \frac{t}{\tau_q}; \ \frac{{\bfchi}\left(t, {\bf X} \right)}{\ell_1}, \frac{{\bfchi}\left(t, {\bf X} \right)}{\ell_2},...,\frac{{\bfchi}\left(t, {\bf X} \right)}{\ell_q} \right), 
\end{array}
\label{traj}
\eea
is expected to be much more rapid and irregular with respect to velocity field fluctuations.

\bigskip

\subsection{Trajectories bifurcations in physical space. Kinematic bifurcation rate}

One point of the physical space is of bifurcation for a fluid particle trajectory if ${\nabla_{\bf x} {\bf u}}(t, {\bf x})$ has at least an eigenvalue with zero real part, and this occurs when its determinant vanishes, i.e. 
\bea
\ds \Sigma_K: \ {\cal D}_K \equiv \det \left( \nabla_{\bf x} {\bf u} (t, {\bf x}) \right) = 0.
\label{det Grad v}
\eea
Such bifurcations, here called kinematic bifurcations or trajectories bifurcations, directly cause the divergence between contiguous trajectories in the physical space. 
Equation (\ref{det Grad v}) defines the surface $\Sigma_K$ in the physical space. Due to  analytical structure (\ref{u l}), which includes many arguments, $\Sigma_K$ 
is expected to be a non--smooth surface which moves in physical space.

Along one particle trajectory, the kinematic bifurcations happen with a rate, $S_K$, defined by
\bea
\begin{array}{l@{\hspace{-0.cm}}l}
\ds S_K = \lim_{T \rightarrow \infty} \frac{1}{T} \int_0^T \delta({\cal D}_K) \vert \frac{D {\cal D}_K}{Dt} \vert \ dt, \\\\
\ds \mbox{where} \ \frac{D \circ}{Dt} = \frac{\partial \circ }{\partial t} 
+ \nabla_{\bf x} \circ \cdot {\bf u} = 
\Sum_k \left( \frac{\partial \circ }{\partial \hat{t}_k} \frac{1}{\tau_k} + \frac{\partial \circ }{\partial \hat{\bf x}_k}  \cdot \frac{\bf u}{\ell_k}\right), \\\\
\ds \hat{t}_k=\frac{t}{\tau_k}, \ \  \hat{\bf x}_k = \frac{\bf x}{\ell_ k}
\end{array}
\label{kin rate}
\eea
and
\bea
\begin{array}{l@{\hspace{-0.cm}}l}
\ds \frac{D {\cal D}_K}{Dt} = {\cal D}_K \ \mbox{tr}\left( \left( \nabla_{\bf x} {\bf u}\right)^{-1}  \frac{D \nabla_{\bf x} {\bf u}}{Dt} \right), \\\\ 
\ds \vert \frac{\partial {\cal D}_K }{\partial \hat{\bf x}_k} \vert \approx \vert \frac{\partial {\cal D}_K }{\partial \hat{\bf x}_h}\vert, \ \forall h, k,
\end{array}
\eea
Specifically, $S_K$ gives the frequency at which a fluid particle trajectory intersect $\Sigma_K$.
Now, from Eq. (\ref{Taylor kolmogorov}) and taking into account that $R_{T min} \approx$10 \cite{Batchelor53}, we expect that
\bea
\begin{array}{l@{\hspace{-0.cm}}l}
\ds \vert \frac{\partial {\cal D}_K  }{\partial t} \vert <<< 
\vert \nabla_{\bf x} {\cal D}_K  \cdot {\bf u} \vert
\end{array}
\label{DK}
\eea
Thus, through Eq. (\ref{P S_NS}), we deduce that, in fully developed turbulence, $S_K>>>S_{NS}$.
In fact, taking into account Eqs. (\ref{kin rate})--(\ref{DK}) and that $\tau_k$ and $\ell_k$ are both continuously distributed, $S_K$ is of the order
\bea
\ds S_K \backsim  \frac{u}{\ell_{min}}
\eea
Therefore, comparing this latter with $S_{NS}$, one obtains
\bea
\begin{array}{l@{\hspace{-0.cm}}l}
\ds \frac{S_K}{S_{NS}} \backsim  R_T^{3/2}  
\end{array}
\label{SK/SNS}
\eea
Now, the minimum value of $R_T$ in homogeneous isotropic turbulence is of the order of 10
(\cite{Batchelor53} and references therein), thus
\bea
\begin{array}{l@{\hspace{-0.cm}}l}
\ds \frac{S_K}{S_{NS}} >>> 1, \\\\ 
\ds \inf\left\lbrace\frac{S_K}{S_{NS}}\right\rbrace  \backsim 40 \ \mbox{for} \ R_T \backsim 10
\end{array}
\label{rates}
\eea 
Following Eqs. (\ref{rates})--(\ref{SK/SNS}), the velocity fluctuations observed along particle trajectories are much more rapid than the fluctuations of velocity field.

\begin{remark}
Unlike the NS--bifurcations that exert a global effect on velocity field, the kinematic bifurcations
give a local influence on the particles trajectories that only acts in close proximity of those points of physical domain which satisfy Eq. (\ref{det Grad v}).
On the contrary, one Navier--Stokes bifurcation, determining a doubling of velocity field, also causes a doubling of $\bfchi$ according to Eq. (\ref{fields}). Therefore, such  doubling of $\bfchi$ is image of the corresponding Navier--Stokes bifurcation in the physical space.  This agrees with the analysis of Ref. \cite{deDivitiis_3}, where the energy cascade is studied by means of a specific bifurcation analysis.
\end{remark}

\bigskip

\section{Statistical Analysis}

The first part of this section deals with the analysis of energy cascade following the Liouville theorem. Such phenomenon is here identified exploiting the equivalence between Eulerian and Lagrangian standpoints.
The statistics of displacement, velocity and temperature fields is then studied through the previous analysis, and an ergodic property is proposed, which is based on fully developed chaos and statistical homogeneity.

\bigskip

\subsection{Equivalence of Euler and Lagrange descriptions. Liouville theorem. Energy cascade.}

The statistical description of motion is given by fluid state variables distribution function, which changes following the Liouville theorem. For each motion description, this theorem is properly formulated through the equations of motion\cite{Nicolis95}. 

The distribution function associated with the Lagrangian standpoint is 
\bea
\ds P_L =P(t, \dot{\bfchi}, \vartheta_m, {\bfchi} )
\label{P_L}
\eea
whereas the PDF relative to Euler point of view can be obtained from Eq. (\ref{P_L}), (\ref{jac}) and (\ref{dot hat chi}), considering the equivalence of the two descriptions, putting  
${\bf x}={\bfchi}(t, {\bf X})$ in $\dot{\bfchi}$ and $\vartheta_m$ with $\hat{\bfchi}= \bfchi$. 
This gives 
\bea
\ds P_E =P(t, {\bf u}, \vartheta, {\bfchi})
\label{P_E}
\eea
Hence, $P$ satisfies the Liouville theorem associated with the motion equations for both the representations. 

In case of Euler representation, the Liouville equation arises from  
Eqs. (\ref{NS_eq Euler})--(\ref{T_eq Euler}) and (\ref{dot hat chi}), and reads as follows \cite{Nicolis95}
\bea
\begin{array}{l@{\hspace{-0.cm}}l}
\ds \frac{\partial P}{\partial t} +
\ds \frac{\delta}{\delta {\bf u}} \cdot \left( P \dot{\bf u} \right) +
\ds \frac{\delta}{\delta \vartheta} \cdot \left( P \dot{\vartheta} \right)  
=0
\end{array}
\label{Liouville Euler}
\eea
where 
$\delta/ \delta {\bf u}$ and $\delta/ \delta {\vartheta}$ 
are functional derivatives with respect to $\bf u$ and $\vartheta$ respectively, 
being $\delta/ \delta \cdot \circ$ the divergence of $\circ$ in the proper functions space.
Second and third term of Eq. (\ref{Liouville Euler}) provide the contribution of stress tensor and heat flux to $\partial P/ \partial t$, and the effect kinetic and thermal energy cascade. 
These latter are here expressed by $\cal H$, being
\bea
\begin{array}{l@{\hspace{-0.cm}}l}
{\cal H} = \ds - \frac{\delta}{\delta {\bf u}} \cdot \left( P \nabla_{\bf x} \ {\bf u}{\bf u}\right)
 - \frac{\delta}{\delta {\vartheta}} \cdot \left( P \nabla_{\bf x} \ {\vartheta}\cdot {\bf u}\right)
\end{array}
\label{H}
\eea 
where first and second terms give, respectively, kinetic and thermal energy cascade contribution to rate of $P$. 

On the other hand, the Liouville equation written in the Lagrangian framework derives form
Eqs. (\ref{NS_eq Lagrange})--(\ref{T_eq Lagrange}) and (\ref{chi}), and is expressed as 
\bea
\begin{array}{l@{\hspace{-0.cm}}l}
\ds \frac{\partial P}{\partial t} +
\ds \frac{\delta}{\delta \dot{\bfchi}} \cdot \left( P \ddot{\bfchi} \right) + 
\ds \frac{\delta}{\delta \vartheta_m} \cdot \left( P \dot{\vartheta}_m \right)  
\ds + \frac{\delta}{\delta {\bfchi}} \cdot \left( P \dot{\bfchi} \right) 
=0
\end{array}
\label{Liouville Lagrange}
\eea
where, the terms appearing in Eq. (\ref{Liouville Euler}) are linked to those of Eq. (\ref{Liouville Lagrange}) through 
\bea
\begin{array}{l@{\hspace{-0.cm}}l}
\ds \frac{\delta}{\delta \dot{\bfchi}}(\circ) = 
 \frac{\delta}{\delta {\bf u}}(\circ) \left(\frac{\delta {\bf u}}{\delta \dot{\bfchi}}\right)
\\\\
\ds \frac{\delta}{\delta {\vartheta}_m}(\circ) =
 \frac{\delta}{\delta {\vartheta}}(\circ)
\left( \frac{\delta {\vartheta}}{\delta {\vartheta}_m}\right)
\end{array}
\label{tr}
\eea
Due to equivalence of the two motion  descriptions, Eq. (\ref{Liouville Lagrange}) is equivalent to Eq. (\ref{Liouville Euler}) if ${\bf x}={\bfchi}(t, {\bf X})$ in Eq. (\ref{tr}). From Eq. (\ref{jac 1}) $\left( \delta {\bf u}/\delta \dot{\bfchi}\right)_{\bfchi}$ and  $\left( \delta {\vartheta}/\delta {\vartheta}_m\right)_{\bfchi}$ are both identity operators, thus combining Eqs. (\ref{Liouville Lagrange}) and (\ref{Liouville Euler}) with Eqs. (\ref{H}), one obtains ${\cal H}$ in terms of derivatives with respect to $\bfchi$, i.e.
\bea
\ds {\cal H} = \frac{\delta}{\delta {\bfchi}} \cdot \left( P \dot{\bfchi} \right) 
\label{H2}
\eea
As $\delta / \delta {\bfchi}\cdot (P \dot{\bfchi})$ expresses the transport of $P$ in physical space and provides both kinetic and thermal energy cascade, these latter consist in a transport phenomenon, where the following vector 
\bea
P \left( 
\begin{array}{l@{\hspace{-0.cm}}c}
\ds  \nabla_{\bf x} {\bf u} {\bf u} \\\\
\ds  \nabla_{\bf x} {\vartheta} \cdot {\bf u} \\\\
\ds  \dot{\bfchi}
\end{array}
\right),
\eea
being a solenoidal field in ${\cal E} \times {\cal X}$, does not modify $\partial P/\partial t$, in particular does not influence neither the kinetic energy rate nor the thermal energy rate.

It is worth to remark that this formulation, Eqs. (\ref{Liouville Euler})--(\ref{Liouville Lagrange}) and (\ref{H2}) hold, in particular, for the temperature. More in general, such analysis applies, without lack of generality, to any passive scalar which exhibits diffusivity.

\bigskip

\subsection{Statistical independence of fluid displacement and velocity field.}

At this stage of the present study, we can say that, in fully developed turbulence, the fluid displacement fluctuations are independent of velocity field.
To justify this, consider now the trajectory of a single particle $\bf X$.
Following Eq. (\ref{rates}), the fluctuations of ${\bfchi}(t, {\bf X})$ are much faster than velocity  field. In detail, during the motion of $\bf X$, the time interval between two contiguous NS--bifurcations will contain a statistically significant number of kinematic bifurcations, expecially when $R_T$ is very high. Hence, the velocity field variations are expected to be substantially irrelevant on the statistics of fluctuations of $\bfchi$, where the distribution 
of the latter does not depend on the particular realization of ${\bf u}(t, {\bf x})$.
Viceversa, the time variations of ${\bfchi}(t, {\bf X})$ does not influence neither ${\bf u}(t, {\bf x})$ nor $\vartheta(t, {\bf x})$ by definition.
Thus the fluctuations of ${\bfchi}$ are supposed to be statistically independent of velocity and temperature fields and this suggests that 
$P(t, {\bf u}, \vartheta, {\bfchi})$ can be factorized as follows
\bea
\begin{array}{l@{\hspace{-0.cm}}l}
\ds P(t, {\bf u}, \vartheta, {\bfchi}) = F(t, {\bf u}, \vartheta) P_{\bfchi}(t, {\bfchi})
\label{hp}
\label{indep stat}
\end{array}
\eea 
Equations (\ref{hp}) and (\ref{H2}) represent important elements of this study.
Specifically, Eq (\ref{hp}) represents the hypothesis of fully developed chaos of this analysis, and Eq. (\ref{H2}) describes the energy cascade phenomenon.
$P_{\bfchi}$, $F$ and $P$ are really functionals of the corresponding fields, defined in the functions spaces $\cal X$, $\cal E$ and ${\cal E}\times {\cal X}$, respectively.
$F$, defining the Eulerian ensemble, provides the statistics of velocity and temperature fields, whereas $P_{\bfchi}$ expresses the statistics of fluid displacement (Lagrangian ensemble).

For what concerns Eq. (\ref{hp}), it is supported by the results of Refs. \cite{Ottino89, Ottino90} (and references therein), where it is observed the that: a)  
$\dot{\bfchi}={\bf u} (t, {\bfchi})$ produce chaotic trajectories also for relatively simple 
mathematical structure of ${\bf u} (t, {\bf x})$ (also for steady fields!). b) 
The flows represented by ${\bf u} (t, {\bfchi})$ stretch and fold continuously 
and rapidly generating a significant level of particles trajectories mixing.

\bigskip

\subsection{Ergodic property.}

Based on previous analysis, an ergodic property relating 
$F$ and $P_{\bfchi}$ is here presented.

Although $\bfchi(t, {\bf X})$ is statistically independent of the statistics of  ${\bf u}(t, {\bf x})$ and $\vartheta(t, {\bf x})$, $P_{\bfchi}$ is linked to $F$. This arises from fully developed chaos, statistical homogeneity and from $S_K/S_{NS}>>$1. 
To formalize this, consider now the following quantity
\bea
\begin{array}{l@{\hspace{-0.cm}}l}
\ds \Upsilon(t, {\bf x}_1, {\bf x}_2,..., {\bf x}_n) = 
\Upsilon\left( {\bf u}(t, {\bf x}_1),...,{\bf u}(t,{\bf x}_p); 
\vartheta(t, {\bf x}_{p+1}),..., \vartheta(t, {\bf x}_n)\right), 
\end{array}
\eea
The average of $\Upsilon$ calculated for ${\bf x}_1 +{\bfchi}$, ..., ${\bf x}_n+{\bfchi}$ as follows
\bea
\begin{array}{l@{\hspace{-0.cm}}l}
\ds \left\langle  \Upsilon_\chi \right\rangle = \lim_{T \rightarrow \infty} \frac{1}{T} \int_0^T \Upsilon_\chi \ dt, \\\\
\ds \Upsilon_\chi \equiv \Upsilon(t, {\bf x}_1+{\bfchi}, ..., {\bf x}_n+{\bfchi}), \\\\
\ds {\bfchi}={\bfchi}(t, {\bf X}), \ \ \forall {\bf X} \in \left\lbrace {\bf X} \right\rbrace,
\end{array}
\eea
can be also expressed in terms of $P_{\bfchi}$ and $F$ by means of the Birkhoff ergodic theorem
\bea
\begin{array}{l@{\hspace{-0.cm}}l}
\ds \left\langle \Upsilon_\chi \right\rangle \equiv \int_{\cal E} \int_{\cal X} P_{\bfchi} F \Upsilon(t, {\bf x}_1+{\bfchi}, ..., {\bf x}_n+{\bfchi}) d {\cal X} d {\cal E}
\end{array}
\label{ave Upsilon}
\eea 
being $\int_{\cal X}$ and  $\int_{\cal E}$ functional integrals, and 
$d {\cal X}$ and $d {\cal E}$ the corresponding elemental volumes in 
the function spaces $\cal X$ and $\cal E$.
Because of fully developed chaos and taking into account that $S_K/S_{NS}$ $>>>$ 1, the average of $\Upsilon_\chi$ calculated over $\cal X$
\bea
\ds \left\langle \Upsilon_\chi \right\rangle_{\cal X} = \int_{\cal X}  P_{\bfchi} \Upsilon(t, {\bf x}_1+{\bfchi}, ..., {\bf x}_n+{\bfchi}) d {\cal X} 
\eea
is supposed to be independent of the specific realization of $\bf u$ and $\vartheta$. This implies that 
$\left\langle \Upsilon_\chi \right\rangle$=$\left\langle \Upsilon_\chi \right\rangle_{\cal X}$.
On the other hand, due to homogeneity, the average of $\Upsilon_\chi$ calculated over $\cal E$,
\bea
\begin{array}{l@{\hspace{-0.cm}}l}
\ds \left\langle \Upsilon_\chi \right\rangle_{\cal E} = \int_{\cal E} F \ \Upsilon(t, {\bf x}_1+{\bfchi},..., {\bf x}_n+{\bfchi})  \ d {\cal E}, 
\end{array}
\eea 
does not depend on $\bfchi$, therefore 
$\left\langle \Upsilon_\chi \right\rangle$=$\left\langle \Upsilon_\chi \right\rangle_{\cal E}$.

Hence, the link between the two ensembles consists in a kind of ergodic property where 
\bea
\begin{array}{l@{\hspace{-0.cm}}l}
\ds \left\langle \Upsilon_\chi \right\rangle_{\cal E} = \left\langle \Upsilon_\chi \right\rangle_{\cal X}, 
\end{array}
\label{ergodicity1}
\eea
While Eq. (\ref{indep stat}) statistically separates Eulerian and Lagrangian ensembles, 
Eq. (\ref{ergodicity1}) gives the link between $F$ and $P_{\bfchi}$ in case of homogeneous turbulence. 

\bigskip

In particular, the statistics of velocity fluctuations along a fluid particle trajectory 
$\dot{\bfchi}\equiv{{\bf u}}(t, {\bfchi})$ is independent of $F$ and follows 
$P_{\bfchi}$. 
In fact, such statistics can be expressed by means of $F$ and $P_{\bfchi}$, 
through the Frobenius--Perron equation
\bea
\ds P_{\dot{\bfchi}}(t, \dot{\bfchi}) = \int_{\cal X} \int_{\cal E} 
F(t, {\bf u}, \vartheta) P_{\bfchi}(t, {\bfchi})
\delta (\dot{\bfchi} - {{\bf u}}(t, {\bfchi}))  d {\cal X} d {\cal E}
\label{FP}
\eea
Due to fully developed turbulence, the integral over $\cal X$ of Eq. (\ref{FP}) is expected to be independent of the particular realization of velocity field, thus $P_{\dot{\bfchi}}(t, \dot{\bfchi})$ is estimated as
\bea
\ds P_{\dot{\bfchi}}(t, \dot{\bfchi}) = \int_{\cal X} P_{\bfchi}(t, {\bfchi})
\delta (\dot{\bfchi} - {\bf {\bf u}}(t, {\bfchi}))  d {\cal X}
\eea
Hence, $\dot{\bfchi}(t, {\bf X})$ and ${\bf u}(t, {\bf x})$ are statistically independent variables related to $P_{\bfchi}$ and $F$, respectively.

\bigskip

\section{Trajectories divergence in fully developed turbulence.}

Due to very frequent kinematic bifurcations and fluid incompressibility, contiguous particles trajectories diverge with each other, exhibit a huge level of chaos showing a continuous relative velocity distribution. The trajectories separation evolution is given by the following equations
\bea
\begin{array}{l@{\hspace{-0.cm}}l}
\ds \dot{{\bfchi} } = {\bf u} (t, {\bfchi}),   \\\\
\ds \dot{{\bfxi} } = {\bf u} (t, {\bfchi}+{\bfxi}) - {\bf u} (t, {\bfchi}),
\end{array}
\label{kin finite}
\eea
where ${\bfchi}(t, {\bf X})$ and  ${\bfchi}(t, {\bf X}')={\bfchi}(t, {\bf X})+{\bfxi}(t, {\bf X}', {\bf X})$ represent two  trajectories associated with the particles $\bf X$ and $\bf X'$, being $\bfxi$ their relative separation vector.
The trajectories separation rate is quantified by the radial velocity component 
calculated for $\vert {\bfxi} \vert=r$ as
\bea
\ds \dot{\xi}_{\xi} =  \frac{\dot{\bfxi} \cdot \bfxi}{\bfxi \cdot \bfxi} \ r
= \left( {\bf u} (t, {\bfchi}+{\bfxi}) - {\bf u} (t, {\bfchi})\right)  \cdot \frac{\bfxi}{\vert \bfxi \vert}
\label{U}
\eea
Now, fluid incompressibility and kinematic bifurcations have significant implications on the interval of variations of $\dot{\xi}_{\xi}$ and on the statistics of the latter. To analyse this, consider now the representation of $\bfxi$ in a proper frame and statistical isotropy
\bea 
\begin{array}{l@{\hspace{-0.cm}}l}
\ds {\bfxi} =   \sum_{k=1}^3 \xi_k {\bf e}_k \equiv \sum_{k=1}^3  \zeta_k \
{\mbox e}^{\varphi_k(t)} {\bf e}_k, \\\\
\ds \varphi_k(0)=0, \ k = 1, 2, 3
\end{array}
\label{xi inc}
\eea
where $E\equiv\left( {\bf e}_1, {\bf e}_2, {\bf e}_3\right)$ is an orthogonal unit vectors system which rotates with respect to the inertial frame $\cal R$ with angular velocity $\bfomega$ depending on the local fluid motion, being $\xi_k \equiv \zeta_k e^{\varphi_k}$ the coordinates of $\bfxi$ in $E$. Specifically, $E$ is chosen in such a way that ${\bf e}_1$ is the direction of the local maximum growth rate --$\varphi_1$-- of $\ln \vert \bfxi \vert$, whereas ${\bf e}_2, {\bf e}_3$ are associated with $\varphi_2$ and $\varphi_3$, respectively. $\zeta_k$=$\zeta_k(t)$ and $\varphi_k$=$\varphi_k(t)$, $k$=1, 2, 3 are slow growing functions of $t$.
The fluid incompressibility provides
\bea 
\begin{array}{l@{\hspace{-0.cm}}l}
\ds \sum_{k=1}^3 \dot{\varphi}_k(t) =0,
\end{array}
\label{xi inc 1}
\eea
Next, the statistical isotropy, gives a condition over $\varphi_k$ compatible with the 
 incompressibility Eq. (\ref{xi inc 1}). This is analytically written as 
\bea
\begin{array}{l@{\hspace{-0.cm}}l}
\ds \varphi_k(t)=\varphi(t) \cos(\beta + \frac{2}{3}\pi (k-1)), \ \ \ k = 1, 2, 3
\end{array}
\label{isotropy}
\eea
As $\bfxi$ is the separation vector between two fluid particles, $\varphi(t)$ is a Lipschitz monotone  differentiable function of $t$ such that 
\bea
\begin{array}{l@{\hspace{-0.cm}}l}
\ds \forall t \in \left(-\infty, \infty \right), \ \ \  0  < \dot{\varphi}(t) \leq L < \infty  \\\\
\ds \lim_{t \rightarrow \pm \infty} \varphi(t) = \pm \infty, \\\\
\ds \lim_{t \rightarrow \pm \infty}  \inf \dot{\varphi} = 0, \\\\
\ds \lim_{t \rightarrow \pm \infty} \sup \dot{\varphi} = L < +\infty,
\end{array}
\label{lip}
\eea
Furthermore, as $\bf e_1$ corresponds to the maximal rising rate direction of $\ln \vert \bfxi \vert$, 
$\varphi_1$ is maximum following Eq. (\ref{isotropy}), and this gives $\beta=0$ for $\varphi>0$, i.e.
\bea
\begin{array}{l@{\hspace{-0.cm}}l}
\ds \varphi_1(t) = \varphi(t), \\\\
\ds  \varphi_2(t)= \varphi_3(t)
= - \frac{\varphi(t)}{2}
\end{array}
\label{xi inc 11}
\eea
Now, to identify the interval of $\dot{\xi}_{\xi}$, this latter is first expressed in terms of $\zeta_k$ and $\varphi_k$ as follows
\bea
\ds \dot{\xi}_{\xi} = \frac{ \dot{\bfxi} \cdot \bfxi}{\bfxi \cdot \bfxi} \ r =
\frac{\ds  \sum_{k=1}^3 \left( \dot{\zeta_k} \zeta_k+ {\zeta_k}^2 \dot{\varphi}_k\right)  e^{2 \varphi_k}} {\ds \sum_{k=1}^3 {\zeta_k}^2 e^{2 \varphi_k}  } \ r
\label{xi lim}
\eea
The interval endpoints of $\dot{\xi}_{\xi}$ are estimated through the limits for $t \rightarrow \pm \infty$ of Eq. (\ref{xi lim}) taking into account that $\varphi$ and $\zeta_k$ are slow growth functions of $t$, being $\varphi$ also monotonic. These are
\bea
\begin{array}{l@{\hspace{-0.cm}}l}
\ds \inf\left\lbrace \dot{\xi}_{\xi} \right\rbrace= 
    \lim_{t\rightarrow - \infty} \inf \dot{\xi}_{\xi} = -\frac{L \ r}{2}, \\\\
\ds \sup\left\lbrace \dot{\xi}_{\xi} \right\rbrace= 
     \lim_{t\rightarrow + \infty} \sup \dot{\xi}_{\xi} = L \ r
\end{array}
\eea
Therefore
\bea
\begin{array}{l@{\hspace{-0.cm}}l}
\ds \dot{\xi}_{\xi} \in \left( -\frac{M}{2},  M \right),   \\\\
\ds  M = r \ L
\end{array}
\label{U range}
\eea
Based on previous analysis, these endpoints are independent of the particular realization of velocity field, and such limits can be also obtained by supposing that the velocity field is frozen at a given instant.

Moreover, from Eq. (\ref{xi inc}), $\bfxi$ tends to align $\bf e_1$, the maximum rising rate direction of $\ln \vert \bfxi \vert$. For this reason, the relative velocity is usefully expressed in the following form 
\bea
\begin{array}{l@{\hspace{-0.cm}}l}
\ds \dot{\bfxi} = \frac{\bfxi_M -\bfxi}{\tau} + {\bfomega} \times {\bfxi} , \\\\ 
\ds {\bfxi}_M = {\bfxi} + \tau  \sum_{k=1}^3{\bf e}_k \zeta_k e^{\varphi_k} \left(\frac{\dot{\zeta_k}}{\zeta_k} + \dot{\varphi}_k \right)   \approx 
\zeta_1 {\bf e_1} {\mbox e}^{\varphi}, \\\\
\ds \tau = \frac{1}{\sup{\dot{\varphi}}} = \frac{1}{L}
\end{array}
\label{align}
\eea 
Accordingly, $\dot{\xi}_{\xi}$ is related to $r$ and $\xi_M$ through the following equation
\bea
\begin{array}{l@{\hspace{-0.cm}}l}
\ds \dot{\xi}_{\xi} -\frac{\xi_M}{\tau}   \cos \alpha  + \frac{r}{\tau} = 0, \\\\
\mbox{where} \ \alpha = \widehat{{\bfxi}{\bfxi}_M},
\end{array}
\label{align 1}
\eea
Equation (\ref{align}) is the extention of the alignment property of the classical Lyapunov vectors presented in \cite{Ott2002} which is here applied to finite--scale separation vectors $\bfxi$
where the variations of this latter are given by $\varphi$, a monotonic Lipschitz function of $t$. This will contribute to describe the turbulent energy cascade, quantifying such phenomenon.

It is worth to remark that the two radial components of velocity difference 
\bea
\begin{array}{l@{\hspace{-0.cm}}l}
\ds \Delta u_r = \left( {\bf u} (t, {\bf x}+{\bf r}) - {\bf u} (t, {\bf x})\right)  \cdot \frac{\bf r}{r}
= u_r' - u_r, \\\\
\ds \dot{\xi}_{\xi} = \left( {\bf u} (t, {\bfchi}+{\bfxi}) - {\bf u} (t, {\bfchi})\right)  \cdot \frac{\bfxi}{r} = u_\xi'-u_\xi
\end{array}
\label{Deltau xi}
\eea
are two different quantities described by two different statistics:
while $\Delta u_r$ varies according to the Navier--Stokes equations, $\dot{\xi}_{\xi}$ changes
following the alignment property of $\bfxi$, therefore $\Delta u_r$ and $\dot{\xi}_{\xi}$ are represented by $F$ and $P_{\bfchi}$, respectively. In other words, $\Delta u_r$ is described by Eulerian ensemble, whereas $\dot{\xi}_{\xi}$ follows the Lagrangian point of view. 
Nevertheless, in isotropic turbulence, the mean square of $\Delta u_r$ and $\dot{\xi}_{\xi}$ are the same. This can be shown taking into account that the
velocity correlation tensor is \cite{Karman38, Batchelor53}
\bea
\ds \left\langle R_{i j} \right\rangle \equiv \left\langle {\bf u} \otimes {\bf u}' \right\rangle_{\cal E} \equiv 
\left\langle u_i u_j' \right\rangle_{\cal E} =
u^2 \left[ -\frac{1}{2 r} f' r_i r_j + \left(f + \frac{r}{2} f' \right) \delta_{i j} \right]
\label{R} 
\eea
and the separation vector $\bfxi$ can be expressed in terms of $\bf r$ through its canonical representation (\ref{xi inc}) 
\bea
\ds {\bfxi} = {\bf Q}{\bf r}
\label{xi r}
\eea
being $\bf Q$ a fluctuating orthogonal matrix giving the orientation of $\bfxi$ with respect to $\cal R$ following Eq. (\ref{xi inc}), $f(t, r)=\langle u_r u_r' \rangle_{\cal E}/u^2$ is the pair correlations of velocity longitudinal components, and $u \equiv \sqrt{\langle {\bf u}\cdot{\bf u}\rangle_{\cal E}/3}$.
Therefore, accounting for isotropic hypothesis, combining Eq. (\ref{ergodicity1}) with $\Upsilon ={\bf u}\cdot{\bf u}'$ and Eqs. (\ref{R}), (\ref{xi r}) and (\ref{Deltau xi}), one obtains
\bea
\begin{array}{l@{\hspace{-0.cm}}l}
\ds \left\langle (\dot{\xi}_{\xi})^2 \right\rangle_{\cal X} = 
\left\langle (\Delta u_r)^2 \right\rangle_{\cal E} \equiv
2 u^2 \left( 1- f \right) 
\end{array}
\label{isotropy_f2}
\eea
This gives the property that, in isotropic turbulence, the relative average kinetic energy between parts of fluid is not influenced by motion description. Although $\dot{\xi}_{\xi}$ and $\Delta u_r$ satisfy Eq. (\ref{isotropy_f2}), these velocity components are distributed following very different distribution functions, in particular their averages are
\bea
\begin{array}{l@{\hspace{-0.cm}}l}
\ds \left\langle \dot{\xi}_{\xi} \right\rangle_{\cal X} > 0, \\\\
\ds \left\langle \Delta u_r \right\rangle_{\cal E} = 0
\end{array}
\label{isotropy_f3}
\eea
where the first equation of (\ref{isotropy_f3}) expresses that contiguous trajectories continuously diverge with each other, whereas the second one establishes that the relative average velocity between two fixed points of space vanishes in homogeneous turbulence.

\bigskip

\section{Statistics of separation rate.}

In this section, the distribution function of $\dot{\xi}_{\xi}$, say $P_{\dot{\xi}_{\xi}}$, 
is achieved by means of the previous analysis. 
To this purpose, we start from the definition of $\dot{\xi}_{\xi}$ 
\bea
\ds \Sigma: \Psi({\bfchi},\dot{\xi}_{\xi} ) \equiv \dot{\xi}_{\xi} - 
\frac{\dot{\bfxi}\cdot {\bfxi}}{{\bfxi}\cdot {\bfxi}} \ r =0
\label{sigmax}
\eea
Equation (\ref{sigmax}) defines a hypersurface $\ds \Sigma \subset \left\lbrace \cal{X} \right\rbrace$ whose measure $\ds m\left(\Sigma \right)$ is independent of $\dot{\xi}_\xi$ due to the hypothesis of fully developed chaos.
In order to obtain $P_{\dot{\xi}_{\xi}}$, observe that 
$\dot{\xi}_{\xi}$ does not depend on $F(t, {\bf u}, \vartheta)$ and its PDF,
related to $P_{\bfchi}$, can be expressed using Frobenius--Perron equation \cite{Nicolis95} and alignment property (\ref{align 1}), i.e.
\bea
\begin{array}{l@{\hspace{-0.cm}}l}
\ds P_{\dot{\xi}_{\xi}} ( \dot{\xi}_{\xi} )
=  
\int_{\cal X} P_{\bfchi} \ \delta\left( \Psi({\bfchi},\dot{\xi}_{\xi} ) \right)   \  d{\bfchi} \\\\ 
\equiv
\ds \int_{\cal X} P_{\bfchi} \ \delta\left( \dot{\xi}_{\xi} -\frac{\xi_M}{\tau}   \cos \alpha  + \frac{\xi}{\tau} \right)  \  d{\bfchi}
\end{array}
\label{FrobeniusPerron}
\eea
where $\delta$ stands for Dirac's delta. Because of homogeneity, $P_{\dot{\xi}_{\xi}}$ does not depend on $\bf x$, and the idea that there are no privileged directions in isotropic turbulence, suggests that $\dot{\xi}_{\xi}$ can be uniformely distributed in its interval.
To proof this, observe that the integral of Eq. (\ref{FrobeniusPerron}) can be expressed as the
layer integral over  $\ds \Sigma$, i.e \cite{Federer69}.
\bea
\ds P_{\dot{\xi}_{\xi}} ( \dot{\xi}_{\xi} )= \int_{\Sigma} \frac{P_{\bfchi}}{\vert {\nabla_{\bfchi} \Psi}\vert} \ d \Sigma
\eea
As $P_{\bfchi}$,  ${\nabla_{\bfchi} \Psi}$ and $m (\Sigma)$ do not depend on ${\dot{\xi}_{\xi}}$,  $P_{\dot{\xi}_{\xi}}$ is constant in the interval of variation of ${\dot{\xi}_{\xi}}$, being
\bea
\ds P_{\dot{\xi}_{\xi}}  ( \dot{\xi}_{\xi} ) = 
\left\lbrace 
\begin{array}{l@{\hspace{-0.cm}}l}
\ds \frac{2}{3}\frac{1}{M}, \ \ \mbox{if} \  \dot{\xi}_{\xi} \in \left( -\frac{M}{2}, M\right)  \\\\
\ds 0 \ \ \mbox{elsewhere} 
\end{array}\right. 
\label{Pl}
\eea
Another equivalent way to proof Eq. (\ref{Pl}) exploits the statistical hypothesis of isotropy
starting from
\bea
\begin{array}{l@{\hspace{-0.cm}}l}
\ds P_{\dot{\xi}_{\xi}} ( \dot{\xi}_{\xi} )
= 
\ds \int_{{\cal X}} P_{\bfchi} \ \delta\left( \dot{\xi}_{\xi} -\frac{\xi_M}{\tau}   \cos \alpha  + \frac{\xi}{\tau} \right)  \  d {\bfchi}
\end{array}
\label{FrobeniusPerron1}
\eea
wherein $\bfxi_M$ is considered to be given.
\begin{figure}
	\centering
	\includegraphics[width=62mm,height=70mm]{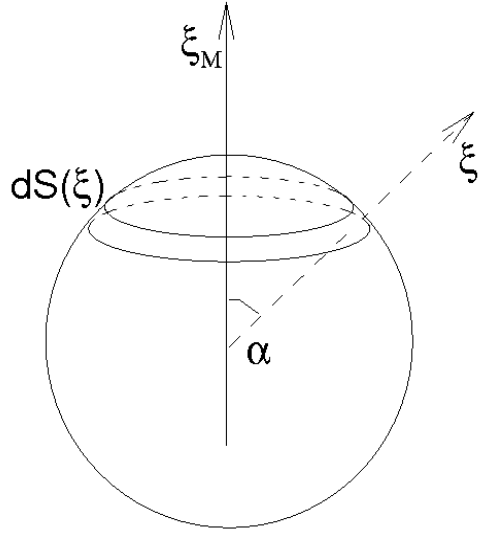}
\caption{Scheme of separation vector and direction of maximum rise of $\bfxi$.}
\label{figura_3}
\end{figure}
Due to isotropy, all the directions of $\bfxi$ are equiprobable, therefore the elemental probability $P_{\bfchi} \ d  {\cal{X}}_1$, calculated near $\Sigma$ ($\Psi=0$), is relative to all those vectors $\bfxi$ such that $\widehat{\bfxi \bfxi_M}\in(\alpha, \alpha+d\alpha)$, thus the corresponding elemental volume is $d  {\cal{X}}_1$ = $\lbrace {\bfchi} \subset {\cal X} \ \vert  \ \widehat{\bfxi \bfxi_M}\in (\alpha, \alpha+d\alpha) \rbrace$. This probability depends on $\alpha$, being proportional to the elemental surface $dS(\xi)$ according to Fig. \ref{figura_3}
\bea
\begin{array}{l@{\hspace{-0.cm}}l}
\ds   P_{\bfchi}  \ d  {\cal{X}}_1 
\ds = \frac{dS(\xi)}{4 \pi \xi^2} = 
\ds \frac{2 \pi \xi^2 \sin \alpha d \alpha}{4 \pi \xi^2} = -\frac{d \cos \alpha}{2}, \ \ \ 
\alpha \in \left( 0, \pi \right),
\end{array}
\label{isotropy 1} 
\eea
Combining Eqs. (\ref{isotropy 1}), (\ref{FrobeniusPerron}) and (\ref{align 1}),
 $P_{\dot{\xi}_{\xi}} ( \dot{\xi}_{\xi} )$ is reduced to be an integral over $\cos \alpha$, where
$\xi_M$ is assigned. This leads to obtain $P_{\dot{\xi}_{\xi}} ( \dot{\xi}_{\xi} )$ in terms of $\dot{\xi}_{\xi}$ 
\bea
\begin{array}{l@{\hspace{-0.cm}}l}
\ds  P_{\dot{\xi}_{\xi}} ( \dot{\xi}_{\xi} ) =  \frac{1}{2} \int_{-1}^1 
\delta\left( -\frac{\dot{\xi}_{\xi} \tau}{\xi_M} - q +\cos \alpha \right)  d \cos \alpha \\\\
\ds = \frac{1}{2} \left( H\left( 1-\frac{\dot{\xi}_{\xi} \tau}{\xi_M} - q\right) -H\left( -1 -\frac{\dot{\xi}_{\xi} \tau}{\xi_M} - q\right) \right), \\\\
\ds q = \frac{r}{\xi_M}
\end{array}
\eea 
where $H$ indicates the Heaveside function. Therefore, $\dot{\xi}_{\xi}$ is uniformely
distributed in the set given by  (\ref{U range}), according to Eq. (\ref{Pl}).

Such PDF gives $\langle \dot{\xi}_{\xi} \rangle_{\cal X} >0$ and provides the link between its statistical moments. In particular, the relation between $\langle \dot{\xi}_{\xi} \rangle_{\cal X}$ and $\langle \dot{\xi}_{\xi}^2 \rangle_{\cal X}$, useful for this analysis, leads to write 
 $\langle \dot{\xi}_{\xi} \rangle_{\cal X}$ in terms of pair velocity correlation according to 
 Eq.(\ref{isotropy_f2})
\bea
\begin{array}{l@{\hspace{-0.cm}}l}
\ds \left\langle \dot{\xi}_{\xi} \right\rangle_{\cal X} =
\frac{1}{2} \sqrt{\left\langle \dot{\xi}_{\xi}^2 \right\rangle_{\cal X}}
\equiv  u \sqrt{\frac{1-f}{2}}
\end{array}
\label{mPl}
\eea 
In conclusion, the PDF (\ref{Pl}) is consequence of the alignment property of $\bfxi$ and of the two following hypotheses: fully developed isotropic turbulence and fluid incompressibility. The first one,
in conjunction with very frequent kinematic bifurcations, generates continuous trajectories divergence, and the combined effect of this latter and fluid incompressibility gives the interval of $\dot{\xi}_{\xi}$: the trajectories divergence, representing instability, generates regions wherein $\dot{\xi}_{\xi}>$0, whereas the incompressibility acts in opposite sense preserving material volume, giving intervals where $\dot{\xi}_{\xi}<$0. As for isotropy, it gives no privileged directions and generates an uniform PDF in the range of $\dot{\xi}_{\xi}$.
These results are in agreement with those observed in Ref. \cite{deDivitiis_6}, where, the author, adopting the Lyapunov theory, shows that the finite scale Lyapunov exponent associated with 
Eq. (\ref{kin finite}) results to be uniformely distributed in its interval.

\bigskip

\section{Closure of von K\'arm\'an--Howarth and Corrsin equations}

Here, the closures of correlation equations are proposed by means of the analysis 
seen in the previous sections.
To this purpose, for sake of reader convenience, von K\'arm\'an--Howarth and Corrsin equations are first renewed. These equations, obtained from the Navier--Stokes and heat equations written in two points of space, $\bf x$ and $\bf x'= x + r$, are
\bea
\begin{array}{l@{\hspace{-0.cm}}l}
\ds \frac{\partial f}{\partial t} = 
\ds  \frac{K}{u^2} +
\ds 2 \nu  \left(  \frac{\partial^2 f} {\partial r^2} +
\ds \frac{4}{r} \frac{\partial f}{\partial r}  \right) +\frac{10 \nu}{\lambda_T^2} f, \\\\
\ds \frac{\partial f_\theta}{\partial t} = 
\ds  \frac{G}{\theta^2} +
\ds 2 \kappa  \left(  \frac{\partial^2 f_\theta} {\partial r^2} +
\ds \frac{2}{r} \frac{\partial f_\theta}{\partial r}  \right) +\frac{12 \kappa}{\lambda_\theta^2} f_\theta,
\end{array}
\label{vk-h}
\eea
wherein $f=\langle u_r u_r' \rangle_{\cal E}/u^2$ and $f_\theta = \langle \vartheta  \vartheta' \rangle_{\cal E}/\theta^2$ are correlations of velocity radial components and of temperature, 
$u \equiv \sqrt{\langle u_r^2 \rangle_{\cal E}}$,  $\theta \equiv \sqrt{\langle \vartheta^2 \rangle_{\cal E}}$, being $\lambda_T \equiv \sqrt{-1/f''(0)}$ and $\lambda_\theta \equiv \sqrt{-2/f_\theta''(0)}$ Taylor and Corrsin microscales, respectively.
$K$ and $G$, providing the turbulent energy cascade, are linked to $k$ and $m^*$, where the latter are, respectively, longitudinal triple velocity correlation and triple correlation between $u_r$ and $\vartheta$, i.e.
\bea
\begin{array}{l@{\hspace{+0.0cm}}l}
\ds K(r) = u^3 \left( \frac{\partial }{\partial r} + \frac{4}{r} \right) 
k(r), 
\ \ \mbox{where} \ \ 
\ds k(r) = \frac{\langle u_r^2 u_r' \rangle_{\cal E}}{u^3}, \\\\
\ds G(r) = 2 u \theta^2 \left( \frac{\partial }{\partial r} + \frac{2}{r} 
\right) m^*(r), 
\ \ \mbox{where} \ \ 
\ds m^*(r) = \frac{\langle u_r \vartheta \vartheta' \rangle_{\cal E}}{\theta^2 u},
\end{array}
\eea
Equations (\ref{vk-h}) are closed if $K$ and $G$ are both in terms of $f$ and $f_\theta$.
Now, if such correlations were calculated following the classical approaches \cite{Karman38, Corrsin_1, Corrsin_2} as averages over $\cal E$ (i.e. through $F(t, {\bf u}, \vartheta)$), $K$ and $G$ will remain unknown quantities unless we assume particular hypotheses about their analytical structures \cite{Karman38, Corrsin_1, Corrsin_2}.
Here, to obtain $K$ and $G$, the correlation equations (\ref{vk-h}) are first formally obtained by means of the Liouville theorem, then $K$ and $G$ are properly identified through the equivalence between Eulerian and Lagrangian motion descriptions. 
For this purpose, von K\'arm\'an--Howarth equation and Corrsin equation are here achieved by multiplying the Liouville equation (\ref{Liouville Euler}) by $u_r u_r'$ and $\vartheta \vartheta'$, respectively, and integrating the so obtained equations over 
$\cal E$ $\times$ $\cal X$, i.e. 
\bea
\begin{array}{l@{\hspace{-0.cm}}l}
\ds \int_{\cal E} \int_{\cal X} u_r u_r' \left( \frac{\partial P}{\partial t} +
\ds \frac{\delta}{\delta {\bf u}} \cdot \left( P \dot{\bf u} \right) +
\ds \frac{\delta}{\delta \vartheta} \cdot \left( P \dot{\vartheta} \right) 
\right) \ d {\cal X} d {\cal E}
=0, \\\\
\ds \int_{\cal E} \int_{\cal X} \vartheta \vartheta' \left( \frac{\partial P}{\partial t} +
\ds \frac{\delta}{\delta {\bf u}} \cdot \left( P \dot{\bf u} \right) +
\ds \frac{\delta}{\delta \vartheta} \cdot \left( P \dot{\vartheta} \right)  
\right) \ d {\cal X} d {\cal E}
=0,
\end{array}
\label{new}
\eea
where second and third addends incorporate $\cal H$ which provides the turbulent energy cascade.
The integrals of $\cal H$ arising from Eqs. (\ref{new}) represent transport terms that do not modify the rates of kinetic and thermal energies \cite{Batchelor53, Corrsin_1}, and identify $K$ and $G$ according to
\bea
\begin{array}{l@{\hspace{+0.0cm}}l}
\ds K = -\int_{\cal E} \int_{\cal X} {\cal H}  \ u_r u_r' \ d {\cal X} d {\cal E} =
-\int_{\cal E} \int_{\cal X} \frac{\delta}{\delta {\bfchi}}\cdot \left( F \ P_{\bfchi} \ \dot{\bfchi}\right)  \ u_r u_r' \ d {\cal X} d {\cal E}, \\\\
\ds G = -\int_{\cal E} \int_{\cal X} {\cal H}  \ \vartheta \vartheta' \ d {\cal X} d {\cal E} =
-\int_{\cal E} \int_{\cal X} \frac{\delta}{\delta {\bfchi}}\cdot \left( F \ P_{\bfchi} \ \dot{\bfchi}\right) \ \vartheta \vartheta' \ d {\cal X} d {\cal E},
\end{array}
\label{closures}
\eea
As for the remaining terms of Eqs. (\ref{new}), these identify the other ones appearing in 
Eqs. (\ref{vk-h}).
Integrating by parts Eqs. (\ref{closures}) with respect to $\bfchi$ and taking into account that $P$=$0$, $\forall {\bfchi} \in \partial {\cal X}$, $K$ and $G$ are so reduced
\bea
\begin{array}{l@{\hspace{+0.0cm}}l}
\ds K = \int_{\cal E} \int_{\cal X} F \ P_{\bfchi} 
\left( \frac{\partial u_r u_r'}{\partial {\bf x}}\cdot \dot{\bfchi}+
       \frac{\partial u_r u_r'}{\partial {\bf x}'}\cdot \dot{\bfchi}' \right)
 \ d {\cal X} d {\cal E}, \\\\
\ds G = \int_{\cal E} \int_{\cal X} F \ P_{\bfchi} 
\left( \frac{\partial \vartheta \vartheta'}{\partial {\bf x}}\cdot \dot{\bfchi}+
       \frac{\partial \vartheta \vartheta'}{\partial {\bf x}'}\cdot \dot{\bfchi}' \right)
 \ d {\cal X} d {\cal E},
\end{array}
\eea
where $F$ gives the average of $u_r u_r'$ and $\vartheta \vartheta'$, whereas $P_{\bfchi}$
statistically describes $\dot{\bfchi}$. Next, in homogeneous isotropic turbulence, the following equations hold \cite{Karman38, Corrsin_1, Corrsin_2}
\bea
\ds \frac{\partial}{\partial {\bf x}'} \left\langle \circ \right\rangle_{\cal E}= -\frac{\partial}{\partial {\bf x}} \left\langle \circ \right\rangle_{\cal E}= \frac{\partial}{\partial {\bfxi}} \left\langle \circ \right\rangle_{\cal E} = \frac{\partial}{\partial r} \left\langle \circ \right\rangle_{\cal E} \frac{\bfxi}{\xi}
\eea
where $\circ = u_r u_r'$, $\vartheta \vartheta'$, and $\dot{\bfxi}$ = $\dot{\bfchi}'$--$\dot{\bfchi}$, being 
${\bf x}'={\bfchi}(t, {\bf X}')$, ${\bf x}={\bfchi}(t, {\bf X})$,
$\dot{\bfchi}'\equiv \dot{\bfchi}(t, {\bf X}')$, $\dot{\bfchi}\equiv \dot{\bfchi}(t, {\bf X})$,
 therefore $K$ and $G$ read as
\bea
\begin{array}{l@{\hspace{+0.0cm}}l}
\ds K  =
u^2 \frac{\partial f }{\partial r}  \left\langle \dot{\xi}_{\xi} \right\rangle_{\cal X}, \\\\
\ds G = \theta^2
\frac{\partial f_\theta}{\partial r} \left\langle \dot{\xi}_{\xi} \right\rangle_{\cal X},
\end{array}
\eea
where $\left\langle \dot{\xi}_{\xi} \right\rangle_{\cal X}$ is linked to $f$ by means of Eq. (\ref{mPl}). This leads to the following closure formulas
\bea
\begin{array}{l@{\hspace{+0.0cm}}l}
\ds K(r) = u^3 \sqrt{\frac{1-f}{2}} \frac{\partial f}{\partial r}, \\\\
\ds G(r) = u \theta^2 \sqrt{\frac{1-f}{2}} \frac{\partial f_\theta}{\partial r},
\end{array}
\label{K}
\label{K closure}
\eea
These equations do not exhibit second order derivatives of autocorrelations, thus Eqs. (\ref{K}) are not closures of diffusive type. These are the result of contiguous trajectories divergence in fully developed chaos. Following Eqs. (\ref{K}), the energy cascade is a sort of propagation phenomenon along $r$ which happens with a propagation speed  $\ds \left\langle \dot{\xi}_\xi\right\rangle_{\cal X}$ depending on $r$ and $u$.
The main asset of Eqs. (\ref{K}) with respect to the other closures is that such equations are not the result of phenomenological assumptions. These are achieved through the equivalence of Lagrangian and Eulerian points of view and the statistical independence of $\bfxi$ and $\bf u$.
This latter allows to analytically express $K$ and $G$ separating the effects of the trajectories divergence (Lagrangian element) from velocity field fluctuations (Eulerian element). 
Due to their theoretical foundation, Eqs. (\ref{K closure}) do not exhibit free model parameters which have to be identified.

These closures coincide with those just obtained by the author in the previous works \cite{deDivitiis_1, deDivitiis_4, deDivitiis_5, deDivitiis_8}, where the formulas are achieved using, among the other things, the finite--scale Lyapunov analysis. 
Here, unlike  such articles, Eqs. (\ref{K closure}) are obtained only 
using the equivalence between Lagrangian and Eulerian motion 
representations, the bifurcations rates and the hypothesis of fully 
developed chaos. 

The novelty of this work with respect to \cite{deDivitiis_1, deDivitiis_4, deDivitiis_5, deDivitiis_8, deDivitiis_9} consists in to identify the energy cascade by means of the equivalence between the two motion representations, and in the estimation of ratio (NS--bifurcations rate)/(kinematic bifurcation rate) which leads to the statistical independence of velocity field and fluid displacement.
Further elements of novelty are the statistical analysis of separation rate and the proposed
 ergodic property, two elements which lead to Eqs. (\ref{K}). 

As regards the results obtained with Eqs. (\ref{K}), the reader is referred to the data given in  \cite{deDivitiis_1, deDivitiis_2, deDivitiis_4, deDivitiis_5, deDivitiis_8}.
In brief, we recall that Refs. \cite{deDivitiis_1,  deDivitiis_4, deDivitiis_5} and \cite{deDivitiis_2} show that Eqs. (\ref{K}) adequately describe the energy cascade phenomenon, reproducing negative values of skewness of velocity and temperature difference 
\bea
\begin{array}{l@{\hspace{+0.0cm}}l}
\ds H_{u 3}(r) \equiv 
\frac{\langle (\Delta u_r)^3 \rangle }{\langle (\Delta u_r)^2 \rangle^{3/2}} 
=
\frac{6 k(r)}{(2(1-f(r)))^{3/2}} \\\\
\ds H_{\theta 3}(r) \equiv 
\frac{\langle (\Delta \vartheta)^2 \Delta u_r \rangle }
{\langle (\Delta \vartheta)^2 \rangle {\langle (\Delta u_r)^2 \rangle}^{1/2}}=
\frac{4 m^*}{ 2(1-f_\theta(r)) (2(1-f(r)))^{1/2}} 
\end{array}
\label{H3}
\eea
and in particular
\bea
\begin{array}{l@{\hspace{+0.0cm}}l}
\ds H_{u 3}(0) = \lim_{r \rightarrow 0} H_{u 3}(r) = - \frac{3}{7}, \\\\
\ds H_{\theta 3}(0) = \lim_{r \rightarrow 0} H_{\theta 3}(r) = - \frac{1}{5}, 
\end{array}
\eea
in agreement with the litarature \cite{Chen92, Orszag72, Panda89, Anderson99, Carati95, Kang2003}, the Kolmogorov law and temperature spectra in line with the theoretical argumentation of Kolmogorov, Obukhov--Corrsin and Batchelor \cite{Batchelor_2, Batchelor_3, Obukhov}, with experimental results \cite{Gibson, Mydlarski}, and with numerical data \cite{Rogallo, Donzis}.
Furthermore, Ref \cite{deDivitiis_8} shows that the proposed closure formulas give a Kolmogorov constant of about 2, and produce correlations self-–similarity in proper interval of $r$, directly caused by the continuous fluid particles trajectories divergence.

\begin{figure}[h]
\centering
\vspace{-0.mm}
\hspace{-45.0mm}
\includegraphics[width=10.cm, height=10.0cm]{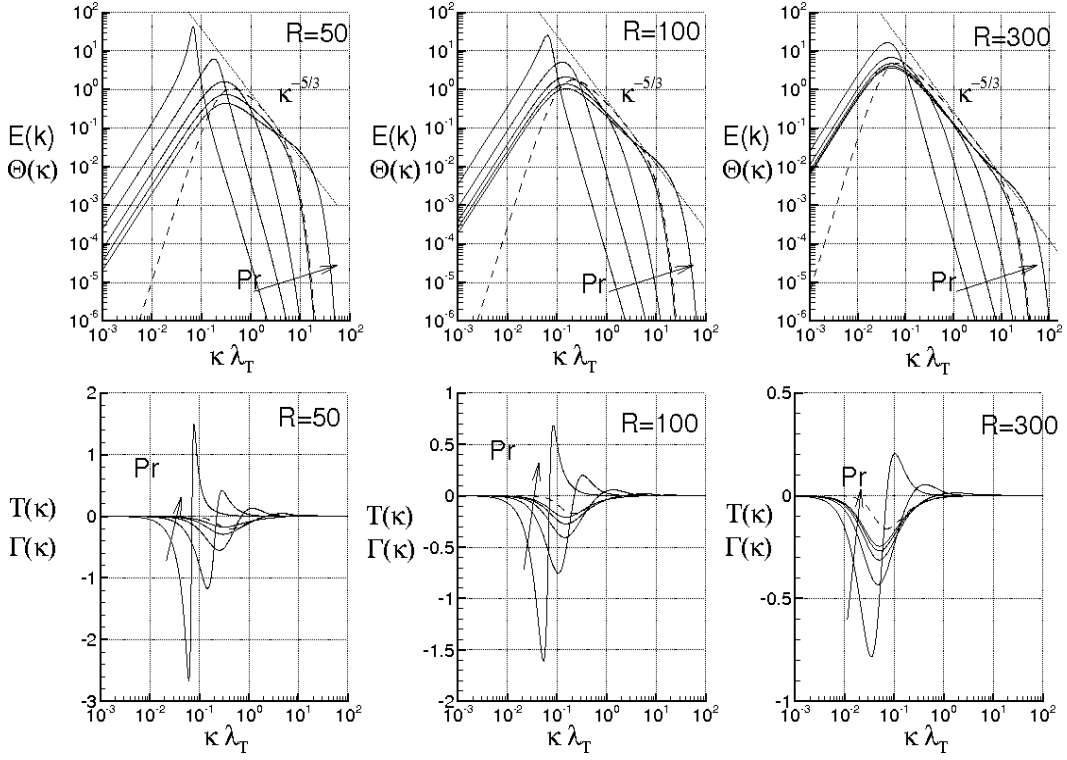}  
\caption{
Spectra for Pr= 10$^{-3}$, 10$^{-2}$, 0.1, 1.0 and 10, at different Reynolds numbers. 
Top: kinetic energy spectrum $E(\kappa)$ (dashed line) and temperature spectra 
$\Theta(\kappa)$ (solid lines). Bottom: velocity transfer function $T(\kappa)$ (dashed line) and temperature transfer function $\Gamma(\kappa)$ (solid line).
}
\label{figura_r3}
\end{figure}

{ \subsection{Results} 
For sake of reader convenience, some of the results just obtained in  \cite{deDivitiis_4, deDivitiis_2} are here reported using the closures (\ref{K}) and self--similarity expressed by $f=f(r/\lambda_T)$, $f_\theta=f_\theta(r/\lambda_\theta)$. 
The fully developed solutions of the correlation equations are obtained for $d\lambda_T/dt=d\lambda_\theta/dt=0$. 
Accordingly, von K\'arm\'an--Howarth and Corrsin equations are reduced to be ordinary differential equations with initial conditions 
$f(0)=1$, $f''(0)=-1/\lambda_T^2$ and $f_\theta(0)=1$, $f''_\theta(0)=-2/\lambda_\theta^2$, where the apex indicates here the differentiation with respect to $r$.
The solutions were numerically obtained using a fourth order Runge--Kutta method with adaptive  step size. 

Figs. \ref{figura_r1} and \ref{figura_r2} show fully developed velocity correlations and the relative  spectra $E(\kappa)$, $T(\kappa)$ numerically calculated for $R_T$=100, 200, 300, 400, 500, 600, being
\bea
\left[\begin{array}{c}
\ds E(\kappa) \\\\
\ds T(\kappa)
\end{array}\right]  
= 
 \frac{1}{\pi} 
 \int_0^{\infty} 
\left[\begin{array}{c}
 \ds  u^2 f(r) \\\\
 \ds K(r)
\end{array}\right]  \kappa^2 r^2 
\left( \frac{\sin \kappa r }{\kappa r} - \cos \kappa r  \right) d r 
\label{Ek}
\eea
where the average kinetic energy is the same for each case. 
\begin{figure}[h]
\centering
\vspace{-0.mm}
\hspace{-0.0mm}
\includegraphics[width=6.cm, height=6.cm]{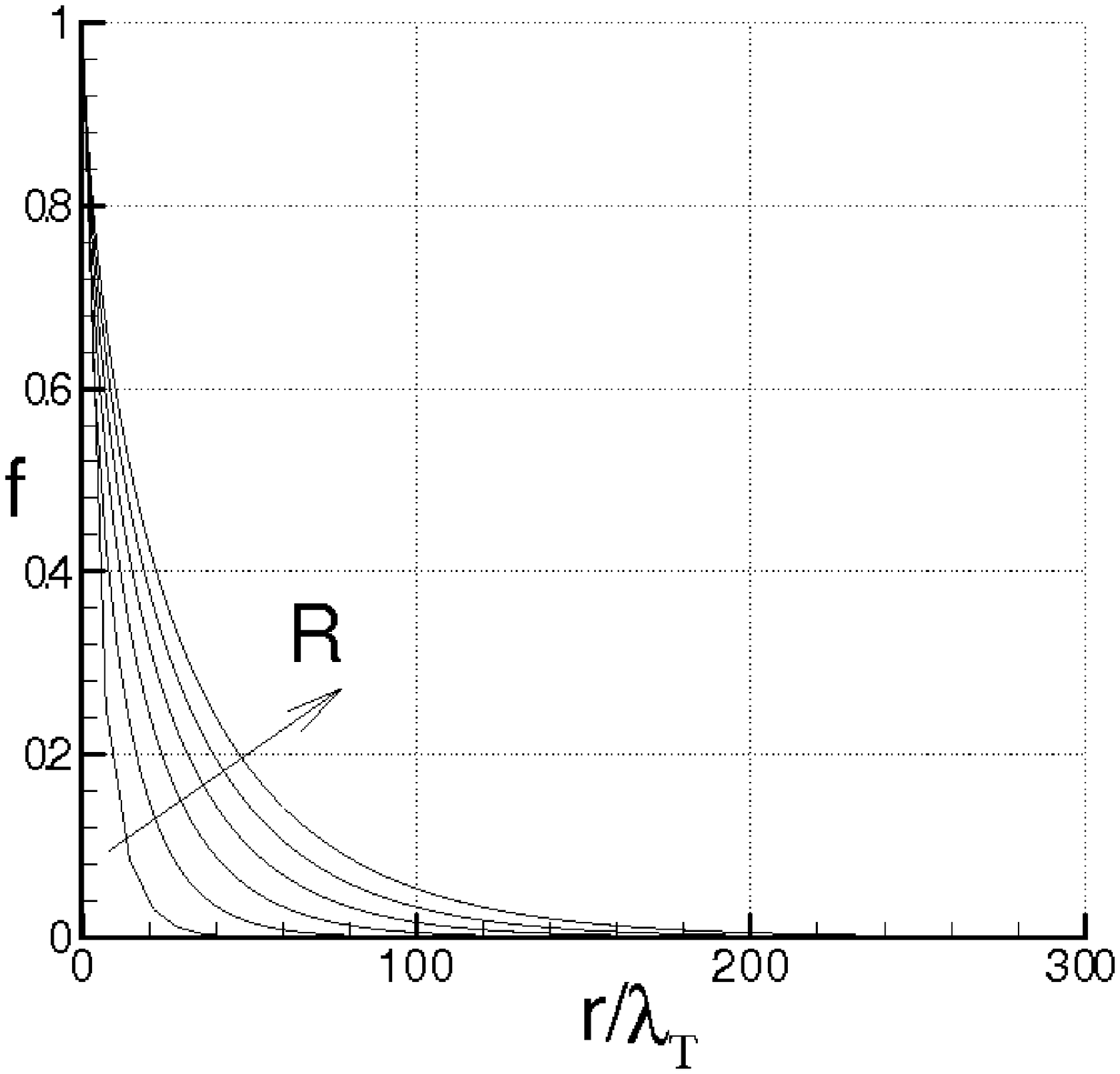} 
\includegraphics[width=6.cm, height=6.cm]{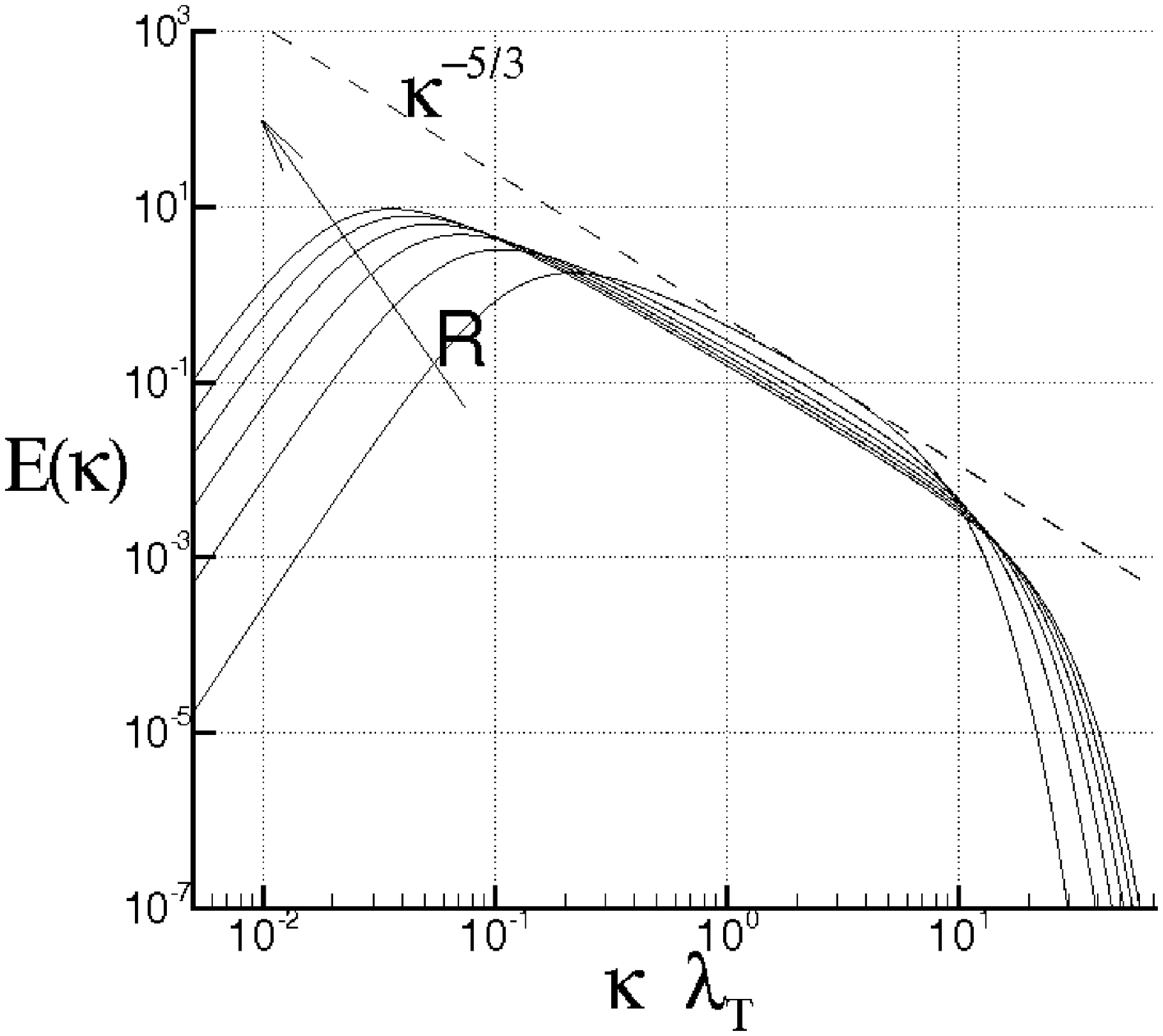}
\caption{ 
Longitudinal velocity correlations (left) and energy spectra (right) at different Taylor scale Reynolds numbers $R_T$=100, 200, 300, 400, 500, 600.
}
\label{figura_r1}
\end{figure}
\begin{figure}[h]
\centering
\vspace{-0.mm}
\hspace{-0.0mm}
\includegraphics[width=6.50cm, height=6.50cm]{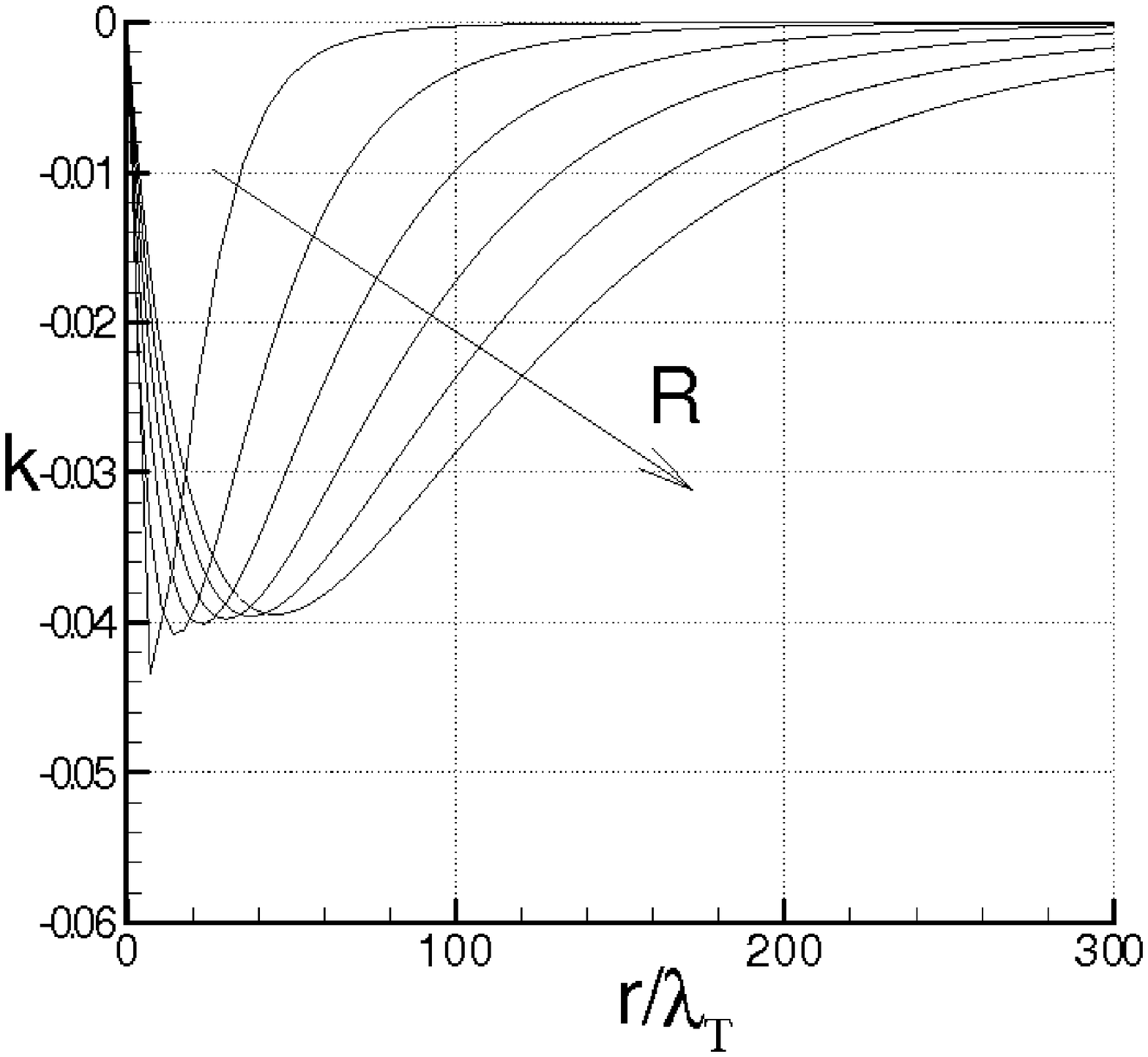} 
\includegraphics[width=6.50cm, height=6.50cm]{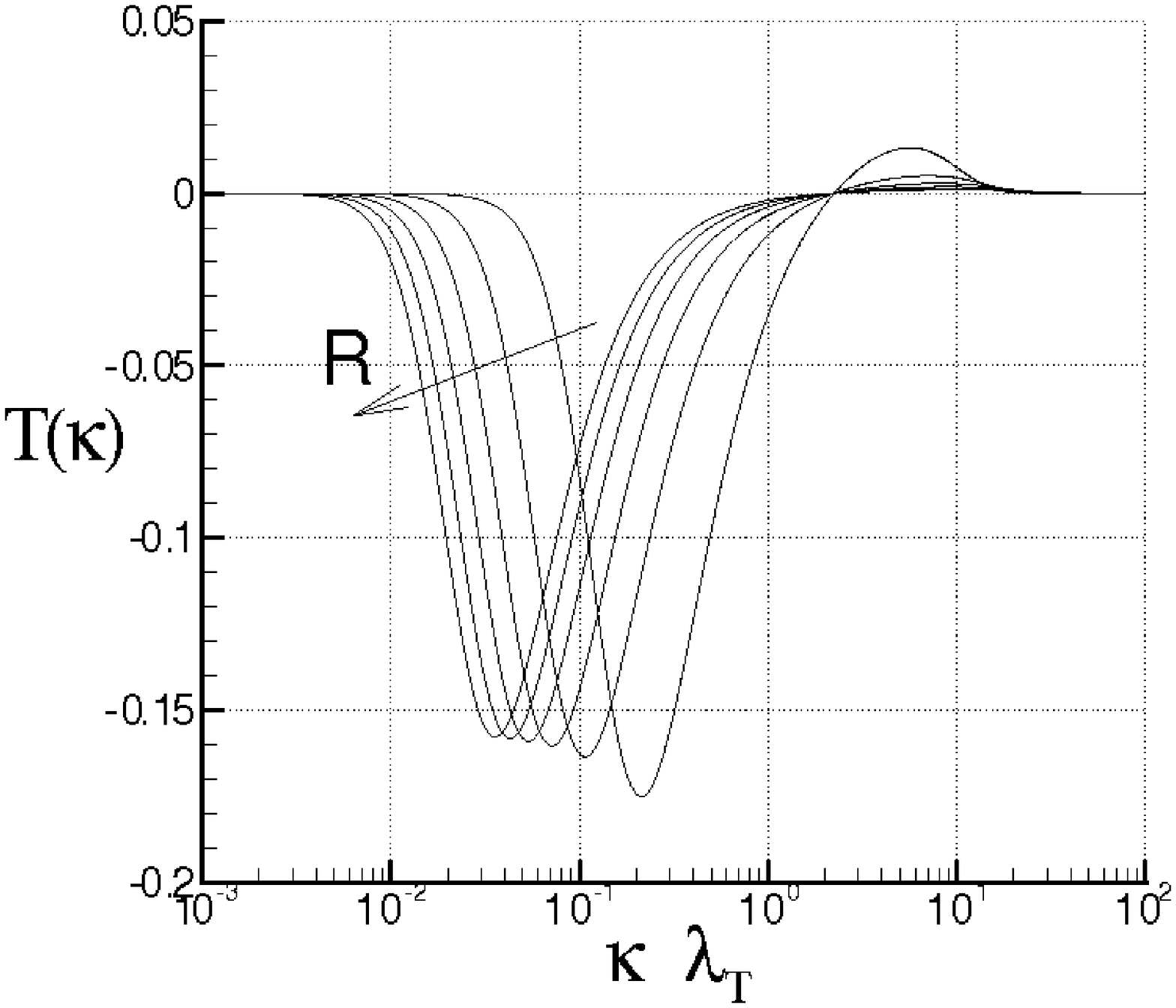}
\caption{ 
Triple longitudinal velocity correlations (left) and the corresponding  spectra (right) at different Taylor scale Reynolds numbers $R_T$=100, 200, 300, 400, 500, 600.
}
\label{figura_r2}
\end{figure}
From such results, the integral  scale of $f$ results to be a rising function of $R_T$, 
$k$ maintains negative and exhibits a minimum of about -0.04 for values of $r/\lambda_T$ which rise with $R_T$. 
$E(\kappa) \approx \kappa^4$ near the origin, then exhibits a maximum and thereafter is about parallel to the dashed line 
$\kappa^{-5/3}$ in a given interval of the wavenumbers. The size of this latter gives the Kolmogorov inertial range and rises as $R_T$ increases. For higher values of $\kappa$, which correspond to scales less than the Kolmogorov length, $E(\kappa)$ decreases much more rapidly than in the inertial range.
$K$ does not modify the kinetic energy, and the proposed closure gives $\int_0^\infty T(\kappa) d \kappa \equiv 0$.

Next, Fig. \ref{figura_r3} provides the temperature spectra $\Theta(\kappa)$ 
and the transfer function $\Gamma(\kappa)$ calculated as follows 
\cite{Ogura}
\bea
\left[\begin{array}{c}
\ds \Theta(\kappa) \\\\
\ds \Gamma(\kappa)
\end{array}\right]  
= 
 \frac{2}{\pi} 
 \int_0^{\infty} 
\left[\begin{array}{c}
 \ds  \theta^2 f_\theta(r) \\\\
 \ds G(r)
\end{array}\right]  
\kappa r \sin \kappa r \ dr 
\label{Tk}
\eea
in such a way that  
\bea
\int_0^\infty \Theta (\kappa)  \ d\kappa = \theta^2, \ \ \
\int_0^\infty \Gamma (\kappa)  \ d\kappa = 0
\label{Tk0}
\eea
$\Theta(\kappa)$ varies with respect to  $R_T$ and $Pr$ in a quite peculiar way.
Specifically, $\Theta(\kappa)$ exhibits different scaling laws $\Theta(\kappa) \approx \kappa^n$ depending on wavenumber interval.
Following the proposed closures, $n \simeq$ 2 as $\kappa \rightarrow$ 0 in any case.
For $Pr=$ 0.001, when $R_T$ ranges from 50 to 300, the temperature spectrum essentially exhibits 
two regions: one in proximity of the origin where $n \simeq 2$, and the other one, at higher values of $\kappa$, 
where $-17/3 < n < -11/3$,  (value very close to $-13/3$). 
The value of $n \approx -13/3$, here obtained in an interval around to 
$\hat{r} \approx$1, is in between the exponent proposed by \cite{Batchelor_3} ($-17/3$) and the value determined by  \cite{Rogallo} ($-11/3$) by means of numerical simulations.
Increasing $\kappa$, $n$ significantly diminishes, and $\Theta(\kappa)$ does not show scaling law.
When $Pr=$0.01, an interval near $\hat{r} \approx 1$ where $-17/3 < n < -13/3$ appears, and this agrees
with \cite{Batchelor_3}.
For $Pr$ =0.1, the previous scaling law vanishes, whereas for $R_T=$ 50 and 100, $n$ changes with $\kappa$, and $\Theta(\kappa)$ does not express clear scaling laws.
When $R_T=300$, the birth of a small region is observed, where $n \approx -5/3$ has an inflection point. For $Pr =$ 0.7 and 1, with $R_T =$ 300, the width of this region is increased, whereas at $Pr$ = 10, and $R_T =$ 300, two regions are observed:
one interval where $n$ has a local minimum with $n \simeq -5/3$, and the other 
one where $n$ exhibits a relative maximum, with $n \simeq -1$.
For larger $\kappa$, $n$ diminishes and the scaling laws disappear.
The presence of the scaling law $n \simeq -5/3$ agrees with the theoretical
arguments of \cite{Corrsin_2, Obukhov} (see also \cite{Mydlarski, Donzis} and references therein).
Figure \ref{figura_r3} also reports (on the bottom) the spectra 
$\Gamma(\kappa)$ (solid lines) and $T(\kappa)$ (dashed lines) which
describe the energy cascade mechanism. 
}

We conclude this study by observing the limits of the proposed closures.
These limits directly derive from the hypotheses under which Eqs. (\ref{K}) are obtained: 
Eqs. (\ref{K}) are valid only in regime of fully developed chaos where the turbulence
exhibit homogeneity and isotropy. Otherwise, during the transition through intermediate stages of turbulence, or in more complex situations with particular boundary conditions, for instance in the presence of wall, Eqs. (\ref{K}) cannot be applied.
In this regard, note that, without the hypotheses of statistical isotropy and homogeneity, the  energy cascade is however identifiable through $\cal H$ and Liouville theorem, whereas the determination of separation rate statistics, closures of correlation equations and other correlations such as pressure--velocity, remains a very difficult task depending on the particular problem. 

\bigskip

\section{Conclusion \label{Conclusion}}

The energy cascade in isotropic homogeneous turbulence is studied through Euler and Lagrange points of view using the bifurcation rates.
The order of magnitude of such rates justifies that velocity field (Eulerian element) and trajectories (Lagrangian element) are statistically uncorrelated, and the turbulent energy cascade is identified by exploiting the equivalence of Eulerian and Lagrangian motion descriptions through Liouville theorem.
A specific ergodic property is then proposed, which relates the statistics of
both the descriptions, and a detailed  analysis of separation rate is presented, which leads to the statistics of radial velocity component. 
Finally, the closures of  von K\'arm\'an--Howarth equation and Corrsin equation are determined. These coincide with those just determined by the author in previous articles \cite{deDivitiis_1,  deDivitiis_4, deDivitiis_5, deDivitiis_8, deDivitiis_9}, corroborating the previous works. These formulas, of nondiffusive nature, allow to adequately describe the phenomenon of  energy cascade.

\bigskip 

\section{Acknowledgments}

This work was partially supported by the Italian Ministry for the Universities 
and Scientific and Technological Research (MIUR). 

\bigskip

\end{document}